\documentclass[onecolumn,aps,unsortedaddress,pre]{revtex4-2}

\usepackage{fullpage}
\usepackage{amsmath, amsfonts}
\usepackage{mathtools}
\usepackage{bm}
\usepackage{dsfont}
\usepackage{subfig}
\usepackage{caption}
\usepackage{setspace}
\usepackage{todo}
\usepackage{natbib}

\usepackage[dvipsnames]{xcolor}

\newcommand{\rhon}{\rho_{\bm n}}
\newcommand{\rhonm}{\rho_{\bm n}^{\bm m}}

\newcommand{\mR}{\mathbb R}

\newcommand{\sF}{\mathcal{F}}

\newcommand{\sH}{\mathcal{H}}

\newcommand{\sI}{\mathcal{I}}%minimizing operator

\newcommand{\bpm}[1]{\begin{pmatrix}#1 \end{pmatrix}}

\newcommand{\bM}{\mathbf{M}}

\begin{document}

\title{\vspace{-2em} An equivalence framework for an age-structured multi-stage representation of the cell cycle}
\author{Joshua C. Kynaston}
    \thanks{Author to whom correspondence should be addressed. Email: josh.c.kynaston@gmail.com}
    \affiliation{Department of Mathematical Sciences, \\University of Bath}
\author{Chris Guiver}
    \affiliation{School of Engineering and The Built Environment, \\ Edinburgh Napier University}
\author{Christian A. Yates}
    \affiliation{Department of Mathematical Sciences, \\ University of Bath}
\begin{abstract}
    We develop theoretical equivalences between stochastic and deterministic models for populations of individual cells stratified by age. Specifically, we develop a hierarchical system of equations describing the full dynamics of an age-structured multi-stage Markov process for approximating cell cycle time distributions. We further demonstrate that the resulting mean behaviour is equivalent, over large timescales, to the classical McKendrick-von Foerster integro-partial differential equation. We conclude by extending this framework to a spatial context, facilitating the modelling of travelling wave phenomena and cell-mediated pattern formation. More generally, this methodology may be extended to myriad reaction-diffusion processes for which the age of individuals is relevant to the dynamics.
\end{abstract}
\maketitle
\section{Introduction}
    Age structure is an important, but often overlooked, element of models for proliferating cells. While a population of cells, for example, might grow according to some Malthusian law asymptotically \cite{bremer_modulation_2008}, this observation gives no information regarding important cellular characteristics such as cell cycle time distributions (CCTDs) or cellular phase durations. Age-structured modelling enjoys application in a wide range of fields, including demography \cite{lindh_age_1999}, cellular migration and invasion \cite{gavagnin_invasion_2019}, targeted therapeutics \cite{gabriel_contribution_2012}, and oncology \cite{iwata_dynamical_2000}.  
    
    The cell cycle is regulated according to various biological processes, many of which depend not only on the present moment in time, but on previous events and properties of the cells. This is inconsistent with the much used \cite{ryser_quantifying_2016,castellanos-moreno_stochastic_2014,baar_stochastic_2016} but incorrect assumption that the cell cycle is a memoryless process. As such, the cell cycle is not easily adapted for simulation via many of the most popular stochastic simulation algorithms for Markov processes, such as the Gillespie direct method \cite{gillespie_exact_1977} or the next reaction method \cite{gibson_efficient_2000}, which rely on the assumption that inter-event times are exponentially distributed. The multi-stage model (MSM) \cite{yates_multi-stage_2017} circumvents this by representing the cell cycle as a sequence of $K$ independent, exponentially distributed stages, which in general need not correspond to the classic phases of the cell cycle.
    
\begin{figure}
        \centering
        \includegraphics[width=8.4cm]{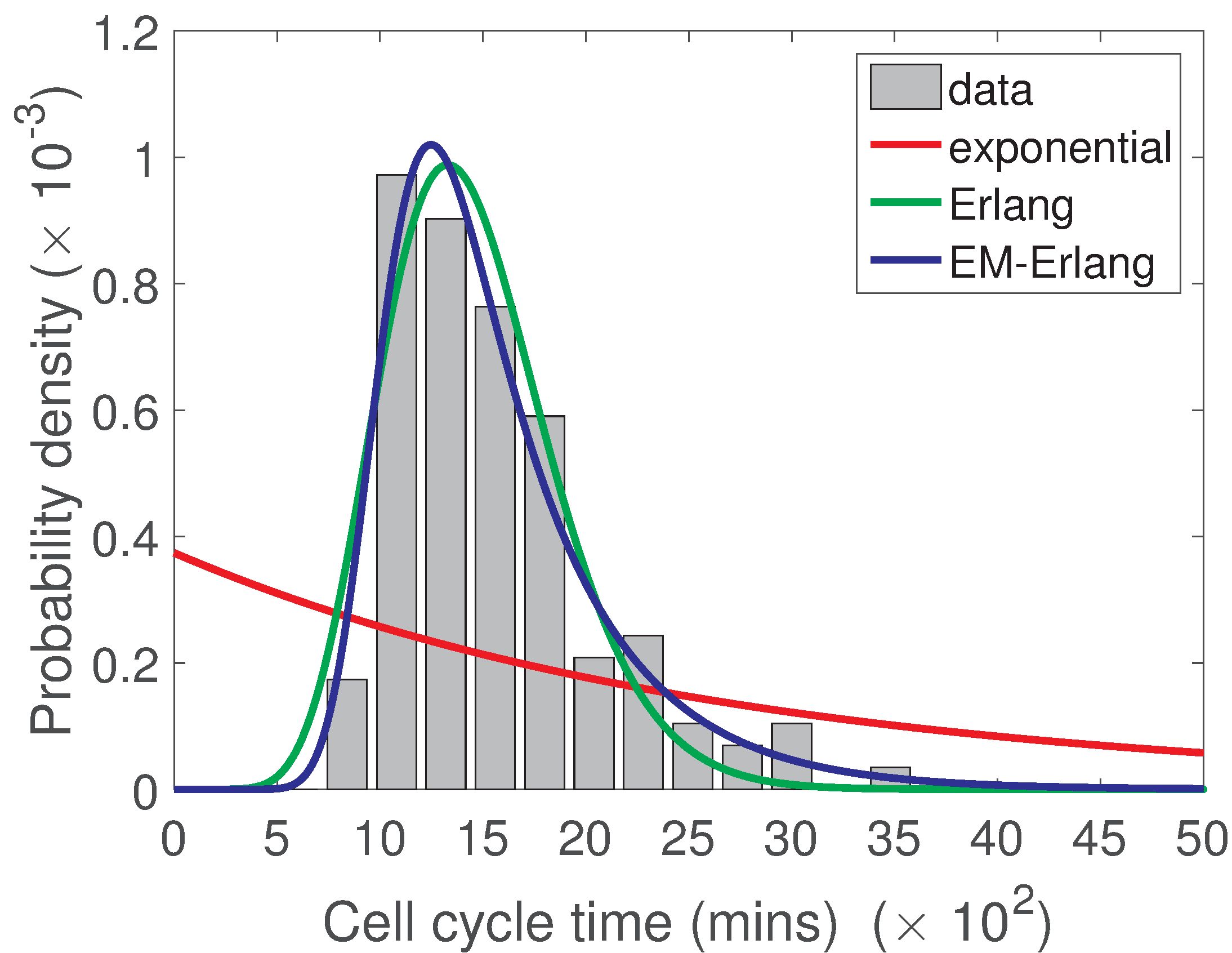}
        \caption{Experimentally determined CCTD in NIH 3T3 mouse embryonic fibroblasts, grown \textit{in vitro} (histogram). Red, green, and blue curves depict the best-fit CCTDs from the exponential, Erlang, and exponentially-modified Erlang distributions, respectively. Reproduced with permission from \cite{yates_multi-stage_2017}.}
        \label{fig:hypoexp-example}
    \end{figure}

    Specifically, the MSM assigns each individual a stage $k$, from which it can transition to stage $k+1$ at a rate $\lambda_{k}$. Cytokinesis is incorporated through the transition of individuals in stage $K$ into two new individuals in stage 1 at a rate $\lambda_K$. The $K$-stage MSM can then be represented by the following reaction system,
    \begin{equation}
        X_1 \xrightarrow{\lambda_1} X_2 \xrightarrow{\lambda_2}\hdots\xrightarrow{\lambda_{K-1}} X_K \xrightarrow{\lambda_K} 2X_1. \label{eq:msm}
    \end{equation}
    The CCTD can then be derived from the convolution of waiting time distributions for each of the $K$ stages. In the simplest case, where all rates $\lambda_k = \lambda$ are equal, the resultant CCTD is the Erlang distribution with shape parameter $K$ and rate parameter $\lambda$. More generally, in the case where the rates $\lambda_k$ are not necessarily equal, the CCTDs of the MSM are of the hypoexponential family; this family is a good fit to many experimentally derived CCTDs \cite{golubev_applications_2016,chao_evidence_2019,cao_analytical_2020,jia_cell_2021}. Figure~\ref{fig:hypoexp-example} compares an experimentally-derived CCTD with best-fit CCTDs from the hypoexponential family. As mentioned, the stages of the MSM do not directly correspond to the classical phases of the cell cycle; however, the choice of the number of stages $K$ and the rate parameters $\lambda_i$ allows for the representation of cell cycle phases as sequences of consecutive stages in the MSM. This grouping of consecutive stages would result in phase length distributions from the hypoexponential family. Hypoexponential distributions have been demonstrated to be a good fit for experimentally derived distributions of phase length \cite{chao_evidence_2019}, motivating the use of the MSM in the present work as a framework through which to analyse the age-structure of cell cycle phases.
    
    One of the most well-known methods for age-structured modelling is the McKendrick-von Foerster equation (MVFE) \cite{mckendrick_applications_1925,von_foerster_remarks_1959}, which takes the form of a linear, first-order integro-partial differential equation describing the evolution of a population's age density. In its canonical form, the density of individuals at any time $t$ with ages in the interval $[a, a+\text d a)$ is defined to be $\rho(a,t)\text da$. It is assumed that individuals may only leave the population via death, which occurs with some rate $\mu(a)$ that depends on the age of the individual. The MVFE is then
    \begin{equation}
        \frac{\partial \rho}{\partial t} + \frac{\partial \rho}{\partial a} = - \mu \rho.
        \label{eq:mvfe}
    \end{equation}
    Further, it is assumed that individuals may enter the population only via birth, which occurs with some rate $\beta(a)$ that depends on the age of the parent individual. This gives rise to the boundary condition, often referred to as the renewal condition,
    \begin{equation}
        \rho(0,t) = \int_0^\infty \beta(\sigma) \rho(\sigma, t)\,\text d\sigma.
        \label{eq:mvfebc}
    \end{equation}
    For the purposes of modelling a cellular population undergoing cytokinesis, one can interpret the death rate $\mu$ to be the rate at which cytokinesis occurs. In this case, one can derive the cytokinesis rate directly from the CCTD; specifically, if $f(a)$ is the probability density function of the CCTD, then
    \[ \mu(a) = \frac{f(a)}{\int_{a}^\infty f(\sigma)\text d\sigma}. \]
    The process of cells entering the population via mitotic divisions is then represented via the renewal condition by setting $\beta(a) = 2\mu(a)$, where any cell which `dies' is replaced by two newborns. The MVFE forms the basis for a wide variety of deterministic age- and/or size-structured models that have seen extensive application in the literature; see, for example \cite{arino_comparison_1993,arino_survey_1997}.
    
    While equations for the evolution of the mean number of cells in each stage of the MSM can be derived \cite{yates_multi-stage_2017}, these equations give no information on the underlying age structure of the system. In Section II, we develop a master equation describing the age and time evolution of the MSM. We then demonstrate that, for large timescales, the mean age distribution of the MSM obeys the MVFE with suitable birth and death rates. Further, we construct the marginal densities from the master equation, from which we derive the deterministic mean-field description of the system; in particular, we derive a system of MVFE-like partial differential equations that describes how the mean density of cells varies with age and time. Similar multi-stage analogues of the MVFE have been described in prior works; for example in \cite{xia_pde_2020}, which proposes a two-stage system of deterministic MVFE-like equations to describe an initiation-adder model for the regulation of bacterial cell size. Further, deterministic mean-field descriptions yielding MVFE-like equations have been derived for birth-death processes \cite{greenman_kinetic_2016}, binary fission-death processes \cite{chou_hierarchical_2016}, and for stochastic sizer-timer models \cite{xia_kinetic_2021} that track both the age- and size-structure of a growing cellular population. None of these works, however, consider the multiphase structure of the cell cycle.
    
    A key motivation for the present work is to provide a foundation for the future development of spatially-extended hybrid methods, extended to incorporate age structure. Spatially-extended hybrid methods are techniques for accurate and efficient simulation of stochastic biological systems that exploit the inherent efficiency of numerical techniques such as finite difference methods, by applying a deterministic description on regions of the domain where stochasticity is not an important driver of the dynamics and coupling them with Markovian simulations in regions where stochasticity dominates. A critical step in the development of hybrid methods is the establishment of an equivalence framework between the underlying stochastic method of interest and any deterministic approximations of said method. 
    
    In Section III, we analyse the mean behaviour of the system over long timescales. Specifically, we demonstrate that when all rates $\lambda_k$ are equal, the normalised mean density converges to a distribution that obeys the steady-state MVFE. Since all transitions and birth events in the MSM are first-order, that is, they are initiated by only a single cell at an age-independent rate, our derived deterministic equations for the mean age distribution of the MSM are exact. In Section IV, we illustrate our equivalence framework numerically via the presentation of two simulated test cases: the first comparing a stochastic simulation of the MSM with the numerically computed solutions to the system of mean density equations, and the second comparing a stochastic simulation of a spatially-extended MSM with the numerically computed solutions to a similarly extended system of mean density equations. In Section V we make some concluding remarks and discuss further avenues of development and application of our theory. Several technical derivations are contained in the Appendix.
    
\section{Dynamics of the age-structured multi-stage model}\label{sec:dynamics}
    In this section, we derive a master equation that describes the full age and time evolution of the age-structured multi-stage model (aMSM). This equation can be used to quantify the stochastic variations of the aMSM exactly. Without this, the variations can only be estimated numerically using stochastic simulation techniques. The technique we employ was first used by Chou and Greenman \cite{chou_hierarchical_2016} to quantify the moments of a simple binary fission process; the present work generalises these techniques to include multiple cellular species (in this case, the stages of the aMSM) undergoing first-order reactions (transitions from one stage to the next) that do not change the age of the cell.
    
    To accomplish this, we make a slight modification to the system of reactions~\eqref{eq:msm}, which ultimately simplifies the boundary conditions of the system. Specifically, we include an additional `stage' that keeps track of twin-pairs in the first stage arising from cytokinesis. Notice that this does not alter the dynamics of the reaction set, as each of a twin-pair of newborn cells are indistinguishable from one another until one advances to the next stage.
    
    \begin{figure}[h!]
    \centering
    \includegraphics[scale=.45]{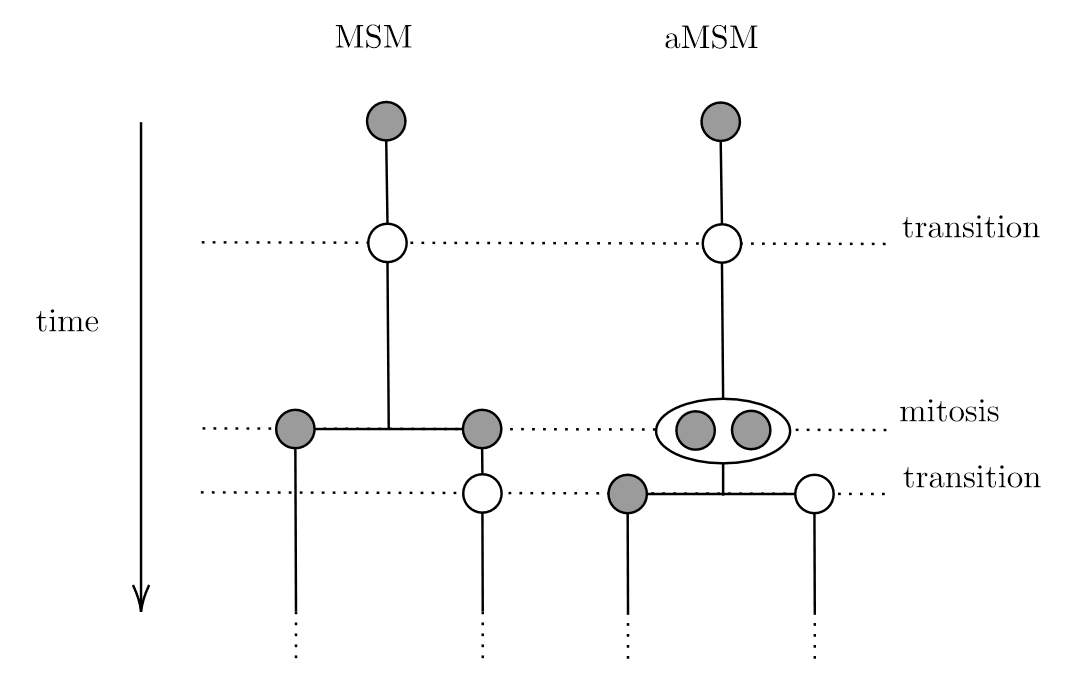}
    \caption{Diagrammatic comparison of the MSM and aMSM with 2 (pseudo)stages, showing a single individual beginning in stage 1 (grey), transitioning to stage 2 (white) before undergoing a cytokinesis event. One of the newborn individuals subsequently transitions to stage 2. Left: Standard MSM. Right: aMSM, with the inclusion of the auxiliary twin-pair stage, represented by two stage 1 individuals enclosed within an ellipse. When one of the twin-pair increments its stage, the twin-pair is removed and replaced by one stage 1 individual and one stage 2 individual.}
    \label{fig:msm_vs_aMSM}
    \end{figure}
    
    To make this precise, suppose that we have a $K$-stage MSM, for a fixed positive integer $K$, where we keep track of the continuous age of each individual. We introduce an auxiliary `stage' consisting of twin-pairs that arise from cytokinesis, which for convenience we will denote with the subscript $K+1$. Fig.~\ref{fig:msm_vs_aMSM} depicts the conceptual difference, for comparison, between the 2-stage MSM and the 2-stage aMSM in the case of a cytokinesis event followed by a stage transition. In contrast to the MSM, the governing reactions of the aMSM are
    \begin{equation}
        Y_1 \xrightarrow{\lambda_1}\hdots\xrightarrow{\lambda_{K-1}} Y_K\xrightarrow{\lambda_K}Y_{K+1}\xrightarrow{2\lambda_1}Y_1 + Y_2.
        \label{eq:aMSM}
    \end{equation}
   Note that the `stages' $Y_k$ in the aMSM, which we call pseudostages, are not all equivalent to the stages $X_k$ in the MSM; in particular, we have $2Y_{K+1} + Y_1 = X_1$ and $Y_k = X_k$ for $k=2,\hdots,K$. Further, a key component of the aMSM is that the age of an individual is set to 0 at the instant of cytokinesis and that subsequent stage transitions do not reset the age clock of an individual to 0. This approach is similar to that used in \cite{chou_hierarchical_2016,greenman_kinetic_2016} in that we extend the dimensionality of the state space to distinguish twin-pairs of individuals through the inclusion of an additional stage $Y_{K+1}$. We additionally adopt the pseudostage nomenclature to distinguish the MSM from the aMSM, the latter of which is identical besides the additional structure endowed by distinguishing twin-pairs.
   
   Assume that at time $t$, for each $k = 1,\hdots,K$, we have $n_k$ individuals in the $k^\text{th}$ pseudostage, and $n_{K+1}$ twin-pairs. Denote by
    \[ \bm n := (n_1,\hdots,n_{K+1}) \]
    the number of individuals in each pseudostage, and by
    \[ \bm a_{n_k} := ((a_k)_1,\hdots,(a_k)_{n_k}) \]
    the vector containing the ages of the individuals in the $k^\text{th}$ pseudostage, where $(a_k)_i$ is the age of the $i^\text{th}$ individual in the $k^\text{th}$ pseudostage. Further, define
    \[ \rho_{\bm n} \left\{\bm a_{n_1}\,;\hdots;\bm a_{n_{K+1}}\,;t\right\}\text d\bm a_{n_1} \hdots \text d \bm a_{n_{K+1}} \]
    to be the probability that at time $t$, after randomly enumerating each individual and twin-pair in the system, the $i^\text{th}$ individual in the $k^\text{th}$ pseudostage has age in $((a_k)_i, (a_k)_i + \text d(a_k)_i)$. Notice that this results in a convenient symmetry property for the probability density function $\rhon$. Namely, the reordering of any age vector $\bm a_{n_k}$ does not change the associated probability. For brevity, define the following operators that modify the index vector $\bm n$,
    \begin{align*}
        \mathcal F_{k} \bm n &= (\hdots, n_k + 1, n_{k+1} - 1,\hdots), &&k = 1,\hdots,K, \\
        \mathcal G \bm n &= (n_1 - 1, n_2 - 1, \hdots, n_{K+1} + 1), &&k = 1,\hdots,K+1, \\
        \mathcal H_k^\pm \bm n &= (\hdots,n_k \pm 1,\hdots), &&k = 1,\hdots,K+1.
    \end{align*}
   We adopt the convention that any unspecified entries remain unchanged under these operators. These operators can be understood intuitively as backward operators on the given state. Specifically, the $\mathcal F_k$ are backward operators corresponding to transition events between pseudostages $k$ and $k+1$, $\mathcal G$ is the backward operator corresponding to the breaking of a twin-pair, and the $\mathcal H^\pm_k$ are the operators corresponding to either the addition or the removal of an individual in the $k$th pseudostage. We will also introduce the notation $\bm a_{n_k}^{(-i)}$ to represent the vector $\bm a_{n_k}$ with the $i^\text{th}$ entry removed. Then, making explicit the assumption that $\rho_{\bm n}$ is differentiable, we see that $\rho_{\bm n}$ obeys the following master equation,
    \begin{align}
        \begin{split}
        \frac{\partial \rho_{\bm n}}{\partial t} + \sum_{k=1}^{K+1} \sum_{i=1}^{n_k} \frac{\partial \rho_{\bm n}}{\partial (a_k)_i} = &\sum_{k=1}^{K} \lambda_k \left ( \frac{n_k + 1}{n_{k+1}} \right) \sum_{i=1}^{n_{k+1}} \rho_{\mathcal F_k \bm n}\left\{\hdots\,; \bm a_{n_k}, (a_{k+1})_i \,; \bm a^{(-i)}_{n_{k+1}}\,; \hdots \,; t \right\} \\
        &+ 2\lambda_1 \left(\frac{n_{K+1} + 1}{n_1 n_2}\right) \sum_{i=1}^{n_1} \sum_{j=1}^{n_2} \rho_{\mathcal G \bm n} \left\{\bm a_{n_1}^{(-i)}\,; \bm a_{n_2}^{(-j)}\,;\, \hdots\,; \bm a_{n_{K+1}}, \omega_{i,j}\, ; t \right\} \\
        &-\left(2 \lambda_1 n_{K+1} + \sum_{k=1}^{K} \lambda_k n_k \right) \rhon, \\
        \end{split}\label{eq:aMSM_master}
    \end{align}
    where $\omega_{i,j} = (a_1)_i = (a_2)_j$ when $(a_1)_i$ and $(a_2)_j$ are equal, and $\omega_{i,j} = \infty$ if not; note that our model makes the implicit assumption that no individuals can have infinite age, and therefore $\omega_{i,j} = \infty$ forces $\rhon = 0$.
    
    The meaning of the terms in equation~\eqref{eq:aMSM_master} can be understood by considering the mechanisms through which one can enter or leave the state $\rhon$. The first term on the right represents the gain in probability from the transition of a pseudostage $k$ individual into pseudostage $k+1$; a transition $Y_k \rightarrow Y_{k+1}$ for $k=1,\hdots,K$, where the transition $Y_{K}\rightarrow Y_{K+1}$ represents a cytokinesis event. The second term represents the gain in probability from the splitting of a twin-pair in pseudostage 1 into both a pseudostage 1 and pseudostage 2 individual; the reaction $Y_{K+1} \rightarrow Y_1 + Y_2$. Finally, the last term on the right represents the loss of probability from leaving the given state via any process in the reaction set; namely, a pseudostage $k$ individual transitioning into pseudostage $k+1$ for $k=1,\hdots,K$, the occurrence of a cytokinesis event, or the splitting of a twin-pair. The boundary conditions can then be stated as follows:
    \begin{align}
    \begin{split}
        \rho_{\bm n}\left\{\hdots\,; \bm a_{n_k - 1}, 0\,;\hdots;t\right\} = \begin{dcases}
        \lambda_K \left( \frac{n_K + 1}{n_{K+1}} \right) \int_0^\infty \rho_{\mathcal F_K \bm n} \left\{\hdots\,;\bm a_{n_K}, \sigma\,;\hdots\,; t \right\}\,\text d\sigma & k = K+1, \\
        0 & k = 1,\hdots, K.
        \end{dcases}
    \end{split}\label{aMSMBC}
    \end{align}
    The first case, $k = K+1$, represents the birth of new age-zero twin-pairs into the first pseudostage via cytokinesis. The remaining boundary conditions reflect the fact that no individuals outside of a twin-pair can have an age of zero.
    
    The marginal densities of $\rho_{\bm n}$ are constructed by integrating out specified ages. To this end, we define the family of marginal densities via a given vector, $\bm m = (m_1,\hdots,m_{K+1})$, as follows
    \begin{equation}
        \rho_{\bm n}^{\bm m} := \int_0^\infty \hdots \int_0^\infty\, \rho_{\bm n}\, \text d \bm a'_{n_1 - m_1} \hdots \text d \bm a'_{n_{K+1} - m_{K+1}},\label{eq:multiple_integral}
    \end{equation}
    where $\bm a'_{n_k - m_k} = ((a_k)_{m_k + 1}, \hdots, (a_k)_{n_k})$ denotes the vector containing the final $n_k - m_k$ entries of $\bm a_{n_k}$. For example, observe that when $\bm m = \bm 0$ is the zero vector, we can interpret $\rho^{\bm 0}_{\bm n}\left\{\hdots;t\right\}$ as the probability that at time $t$, we find $n_k$ individuals in the $k^\text{th}$ pseudostage and $n_{K+1}$ twin-pairs, irrespective of the ages of the individuals. The general equation for the family of marginal densities can then be found by integrating equation~\eqref{eq:aMSM_master} as indicated in equation~\eqref{eq:multiple_integral}, giving
    \begin{align}
    \begin{split}
        \frac{\partial \rhonm}{\partial t} + \sum_{k=1}^{K+1} \sum_{i=1}^{m_k} \frac{\partial \rhonm}{\partial (a_k)_i} &= \lambda_K \frac{(n_K + 1)(n_{K+1} - m_{K+1})}{n_{K+1}} \rho_{\mathcal F_K \bm n}^{\bm m} \\
        &\quad - \left(2 \lambda_1 n_{K+1} + \sum_{k=1}^{K} \lambda_k n_k \right) \rhonm \\
        &\quad + \sum_{k=1}^{K} \lambda_k \frac{(n_k + 1)(n_{k+1} - m_{k+1})}{n_{k+1}}\rho_{\mathcal F_k \bm n}^{\bm m} \\
        &\quad +\sum_{k=1}^{K} \lambda_k \frac{n_k + 1}{m_{k+1}} \sum_{i=1}^{n_{k+1}} \rho_{\mathcal F_k \bm n}^{\mathcal F_k \bm m}\left\{\hdots\,; \bm a_{m_k}, (a_{k+1})_i \,; \bm a^{(-i)}_{m_{k+1}}\,; \hdots \,; t \right\} \\
        &\quad + 2\lambda_1 \frac{(n_{K+1} + 1)(n_1 - m_1)(n_2 - m_2)}{n_1 n_2} \rho_{\mathcal G \bm n}^{\bm m} \\
        &\quad + 2\lambda_1 \frac{(n_{K+1} + 1)(n_2 - m_2)}{n_1 n_2} \sum_{i=1}^{m_1} \rho_{\mathcal G \bm n}^{\mathcal H^{+}_{K+1} \mathcal H^-_1 \bm m} \left\{\bm a_{m_1}^{(-i)}\,;\hdots;t\right\} \\
        &\quad + 2\lambda_1 \frac{(n_{K+1} + 1)(n_1 - m_1)}{n_1 n_2} \sum_{j=1}^{m_2} \rho_{\mathcal G \bm n}^{\mathcal H^{+}_{K+1} \mathcal H^-_2 \bm m} \left\{\hdots\,;\bm a_{m_2}^{(-j)}\,;\hdots;t\right\} \\
        &\quad + 2\lambda_1 \frac{(n_{K+1} + 1)}{n_1 n_2} \sum_{i=1}^{m_1} \sum_{j=1}^{m_2} \rho_{\mathcal G \bm n}^{\mathcal G \bm m} \left\{\bm a_{m_1}^{(-i)}\,; \bm a_{m_2}^{(-j)}\,;\, \hdots\,; \bm a_{m_{K+1}}, \omega_{i,j}\, ; t \right\}.
    \end{split} \label{aMSM_MARG}
    \end{align}
    The boundary conditions are obtained by integrating equation~\eqref{aMSMBC} in the same manner, yielding
    \begin{align}
    \begin{split}
        \rho_{\bm n}^{\bm m} \left\{\bm a_{n_1}\,; \hdots\,; \bm a_{n_k - 1}, 0\,; \hdots\,; \bm a_{n_{K+1}}\,;t\right\} = \begin{dcases}
        \lambda_K \left( \frac{n_K + 1}{n_{K+1}} \right) \rho_{\mathcal F_K \bm n}^{\mathcal H_{K}^- \bm m} & k = K+1, \\
        0 & k = 1,\hdots,K.
        \end{dcases}
    \end{split}\label{aMSMBC_MARG}
    \end{align}
    Full details of the derivation of the system~\eqref{aMSM_MARG}-\eqref{aMSMBC_MARG} can be found in Appendix A. The system~\eqref{aMSM_MARG}-\eqref{aMSMBC_MARG} forms a hierarchy of equations which is likely intractable to solve analytically; however, significant cancellation occurs when one investigates only the mean behaviour of the system. Specifically, let us define the mean age density for the $k^\text{th}$ stage,
    \begin{equation}
        f_k(a,t) := \sum_{\mathbf n} n_k \rhon^{\bm e_k}, \label{mean-age}
    \end{equation}
    for $k = 2,\hdots, K$, where $\bm e_k$ is the $k^\text{th}$ standard unit vector in $\mathbb R^{K+1}$. Since the first stage is made up of both individuals and twin-pairs, we define
    \begin{equation}
        f_1(a,t) := \sum_{\mathbf n} n_1 \rhon^{\bm e_1} + 2\sum_{\mathbf n} n_{K+1} \rhon^{\bm e_{K+1}} \label{mean-age2}
    \end{equation}
   as the mean age density of the first stage. Then, applying the definitions~\eqref{mean-age}-\eqref{mean-age2} to equations~\eqref{aMSM_MARG}-\eqref{aMSMBC_MARG} yields the following system of linear first-order partial differential equations,
    \begin{align}
        \begin{split}
            \frac{\partial f_k}{\partial t} + \frac{\partial f_k}{\partial a} = \begin{cases}
            -\lambda_1 f_1 & k = 1, \\
            -\lambda_k f_k + \lambda_{k-1} f_{k-1} & k = 2,\hdots,K+1,
            \end{cases}
        \end{split}
        \label{eq:aMSM_pde}
    \end{align}
    with boundary conditions
    \begin{align}
        \begin{split}
            f_k(0,t) = \begin{dcases}
            2\lambda_K \int_0^\infty f_K(\sigma, t)\,\text d \sigma & k = 1, \\
            0 & k = 2,\hdots,K+1,
            \end{dcases}
        \end{split}
        \label{eq:aMSM_bc}
    \end{align}
    for $k = 1,\hdots,K$, yielding a multi-stage analogue of \eqref{eq:mvfe}-\eqref{eq:mvfebc}.

\section{Asymptotic age distribution}
    We now turn our attention to the combined age distribution of the aMSM. Throughout this section we will use a caret grapheme to denote \textit{probability} density functions, such as $\hat\zeta$, as opposed to density functions, such as $f_k$. We will also use a superscript asterisk to denote limiting quantities. 
    
    We wish to determine the time-asymptotic behaviour of $\hat\zeta(a,t)$, defined such that the quantity $\hat\zeta(a,t)\text da$ is the probability that at time $t$ we find an individual, in any pseudostage, with age in $[a, a + \text da)$. For simplicity, we will henceforth take $\lambda_k = \lambda$ for $k=1,\hdots,K+1$, for some fixed $\lambda>0$. This will ultimately lead to asymptotic agreement between the combined age distribution of the aMSM and the solution to \eqref{eq:mvfe}-\eqref{eq:mvfebc} with a cytokinesis rate derived from the Erlang distribution, and thereby an Erlang CCTD. We explore this equivalence numerically in Section 4. In generality, different values for the $\lambda_k$ result in a cytokinesis rate derived from the hypoexponential family; however, a full derivation of this is beyond the scope of the present work. 
    
    Define
    \begin{equation}
         M_k(t) := \int_0^\infty f_k(\sigma, t)\,\text d\sigma \label{eq:meaN_cell_counts}
    \end{equation}
    to be the mean number of individuals in the $k^\text{th}$ pseudostage at time $t$, and
    \begin{equation}
        P_k(t) := \frac{M_k(t)}{\sum_{k=1}^K M_k(t)}\label{eq:prop_definition}
    \end{equation}
    to be the mean proportion of individuals in the $k^\text{th}$ stage at time $t$. Note that this can be interpreted as the probability that a randomly selected individual from the population at time $t$ is in the $k^\text{th}$ stage. The combined age distribution $\hat\zeta(a,t)$ is more naturally written in terms of stage-wise age distributions; to this end, we define
    \begin{equation}
        \hat f_k(a,t) := \frac{f_k(a,t)}{\int_0^\infty f_k(\sigma, t)\,\text d\sigma}, \label{eq:stagewise_age_dist}
    \end{equation}
    where $\hat f_k(a,t) \text da$ gives the probability that, at time $t$, the age of a randomly selected individual in the $k^\text{th}$ stage lies in $[a, a + \text da)$. Since the total population can be partitioned according to stage, it follows from the law of total probability that
    \begin{equation}
        \hat\zeta(a,t) = \sum_{k=1}^K P_k(t) \hat f_k(a,t).\label{eq:zeta_sum}
    \end{equation}
    It remains then to characterise the asymptotic evolution of the $\hat f_k$ by applying definition~\eqref{eq:stagewise_age_dist} to~\eqref{eq:aMSM_pde}-\eqref{eq:aMSM_bc}, giving
    \begin{equation*}
        \frac{\partial \hat f_k}{\partial t} + \frac{\partial \hat f_k}{\partial a} = 
            \begin{dcases}
            -\lambda \hat f_1 - \frac{1}{M_1}\frac{\text d M_1}{\text d t} \hat{f}_1 & k = 1, \\
            -\lambda \hat f_1 - \frac{1}{M_k}\frac{\text d M_k}{\text d t} + \lambda \frac{M_{k-1}}{M_k} \hat f_{k-1} & k = 2,\hdots,K,
            \end{dcases}
    \end{equation*}
    with boundary conditions
    \begin{equation*}
        \hat f_k(0,t) = 
        \begin{dcases}
        2\lambda \frac{M_K}{M_1} & k = 1, \\
        0 & k = 2,\hdots,K.
        \end{dcases}
    \end{equation*}
   
    We now consider the behaviour of $\hat f_k$ over large timescales, making the assumption that $\hat f_k(a,t)$ converges to some steady-state distribution as $t \rightarrow \infty$. Specifically, we define
    \begin{equation}
        \hat f^*_k(a) := \lim_{t \rightarrow \infty} \hat f_k(a,t).\label{eq:stage_density_limit_def}
    \end{equation}
    The steady-state behaviour of the population means and proportions can be shown to satisfy the following identities \cite{yates_multi-stage_2017},
    \begin{align}
        \lim_{t \rightarrow \infty} P_k(t) &= (2^{1/N})^{N-k} (2^{1/N} - 1), \label{eq:ss1}\\
        \lim_{t \rightarrow \infty} \frac{M_{k-1}}{M_k} &= 2^{1/N}, \\
        \lim_{t \rightarrow \infty} \frac{M_K}{M_1} &= 2^{1/N - 1},\\
    \intertext{and}
        \lim_{t \rightarrow \infty} \frac{1}{M_k}\frac{\text d M_1}{\text d t} &= \lambda (2^{1/N} - 1). \label{eq:ss2}
    \end{align}
    The equations~\eqref{eq:ss1}-\eqref{eq:ss2}, when combined with the definition of $\hat f^*_k$, yield the following system of ordinary differential equations for the age distribution of each pseudostage,
    \begin{align}
        \frac{\text d \hat f^*_k}{\text d a} &= 
        \begin{dcases}
        -\lambda^* \hat f^*_1 & k = 1, \\
        -\lambda^* \hat f^*_k + \lambda^* \hat f^*_{k-1} & k = 2,\hdots,K,
        \end{dcases} \label{eq:stage_wise_distributions} \\
        \intertext{with boundary conditions}
        \hat f^*_k(0) &= \begin{dcases}
        \lambda^* & k = 1,\\
        0 & k=2,\hdots,K,
        \end{dcases}\label{eq:stage_wise_distributions_IC}
    \end{align}
    where $\lambda^* = \lambda 2^{1/K}$. The system~\eqref{eq:stage_wise_distributions}-\eqref{eq:stage_wise_distributions_IC} can be solved exactly in this case, giving
    \begin{equation}
        \hat f^*_k(a) = \frac{{\lambda^*}^k a^{k-1}e^{\lambda^*a}}{(k-1)!},\label{eq:stage_density_limit}
    \end{equation}
    which is the probability density function of the Erlang distribution with shape parameter $k$ and rate parameter $\lambda^*$. It can then be demonstrated, by inserting expressions~\eqref{eq:stage_density_limit} and \eqref{eq:ss1} into equation~\eqref{eq:zeta_sum}, that
    \begin{equation}
        \hat \zeta^*(a) := \lim_{t\rightarrow\infty}\hat\zeta(a,t) = 2re^{-ra}\int_a^\infty p(\sigma)\,\text d\sigma,
        \label{eq:exact_combined_dist}
    \end{equation}
    where $r := \lambda^* - \lambda$ and $p$ is the probability density function of the Erlang distribution with shape parameter $K$ and rate parameter $\lambda$ and, in particular, the CCTD of the MSM.
    
    The parameter $r$ can be viewed as the intrinsic growth rate of the population. In particular, it can be demonstrated (see Appendix B) that the mean individual counts $M_k(t)$ satisfy
    \begin{equation*}
        \lim_{t\rightarrow\infty} \sum_{k=1}^K M_k(t) e^{-rt} = C,
    \end{equation*}
    for some constant $C$ as $t \rightarrow \infty$. This demonstrates that asymptotically, the mean age distribution of the aMSM converges to some steady-state combined age distribution. Indeed, this limiting behaviour of the aMSM is captured by the MVFE. Specifically, the method of similarity solutions can be used to demonstrate that $\zeta^*(a)$ is a solution to the steady-state MVFE~\eqref{eq:mvfe}-\eqref{eq:mvfebc};
    \begin{align*}
        \frac{\text d \hat \zeta^*}{\text d a} &= -\left(\frac{p}{\int_a^\infty p(\sigma)\text d\sigma}\right)\hat \zeta^*, \\
        \hat \zeta^*(0) &= 2 \int_0^\infty \hat \zeta^*(\sigma) p(\sigma)\,\text d\sigma.
    \end{align*}
    
\section{Simulation results}
    This section contains, for illustrative purposes, a numerical demonstration of our model behaviour as described in Section II through two simulated test cases. We demonstrate that the asymptotic behaviour of the combined age distribution of the deterministic aMSM~\eqref{eq:zeta_sum} agrees with the stochastic aMSM, and heuristically quantify how rapidly the stochastic aMSM converges to its deterministic limit~\eqref{eq:exact_combined_dist}. Further, we showcase the utility of our theory by applying it to a spatial context where cells are permitted to diffuse throughout some finite domain and identify a spatial extension of the MVFE that provides good agreement with the stochastic, spatial aMSM.
    
    \subsection{Uniform-age initial condition}
    In this first test problem, we compare the deterministic densities of the aMSM \eqref{eq:aMSM_pde}-\eqref{eq:aMSM_bc} as approximated via a finite difference method, with the results of a stochastic simulation of the aMSM~\eqref{eq:aMSM} with trajectories generated via the Gillespie algorithm \cite{gillespie_exact_1977}. In our simulations, we take $\alpha_i = i \Delta \alpha$, for $i = 0,\hdots,N$, to be our age discretisation with grid spacing $\Delta\alpha$, on which the numerical solutions are computed and the stochastic trajectories are discretised.
    
    The system~\eqref{eq:aMSM_pde}-\eqref{eq:aMSM_bc}, which we are approximating numerically, admits solutions on the unbounded domain $[0, \infty)\times[0,\infty)$. Since $f_k(a,t)\rightarrow 0$ as $a\rightarrow\infty$ for all $t$, we must therefore select $N$ and $\Delta\alpha$ such that the truncation point $\alpha_N$ yields a vanishingly small $f_k(\alpha,t)$ for all $\alpha>\alpha_N$. We denote our numerical approximation to the system~\eqref{eq:aMSM_pde}-\eqref{eq:aMSM_bc} by $\tilde{f_k}$, and the mean age density of the aMSM as approximated via the Gillespie algorithm by $\tilde{g_k}$. We also denote by
    \begin{equation}
        \tilde{\zeta}(\alpha_i,t) := \frac{\sum_{k=1}^K \tilde g_k(\alpha_i,t)}{\sum_{k=1}^K \sum_{j=0}^N \tilde g_k(\alpha_j,t)}
        \label{eq:approx_combined_dist}
    \end{equation}
    the stochastic approximation to the combined age distribution $\hat\zeta$ as defined in~\eqref{eq:zeta_sum}, with the goal of demonstrating heuristically that
    \begin{equation}
        \lim_{t\rightarrow\infty} \widetilde\zeta (a_i,t) = \hat \zeta^*(a_i).
    \end{equation} 
    
\begin{figure}
        \centering
        \subfloat[]{\includegraphics[width=0.5\textwidth]{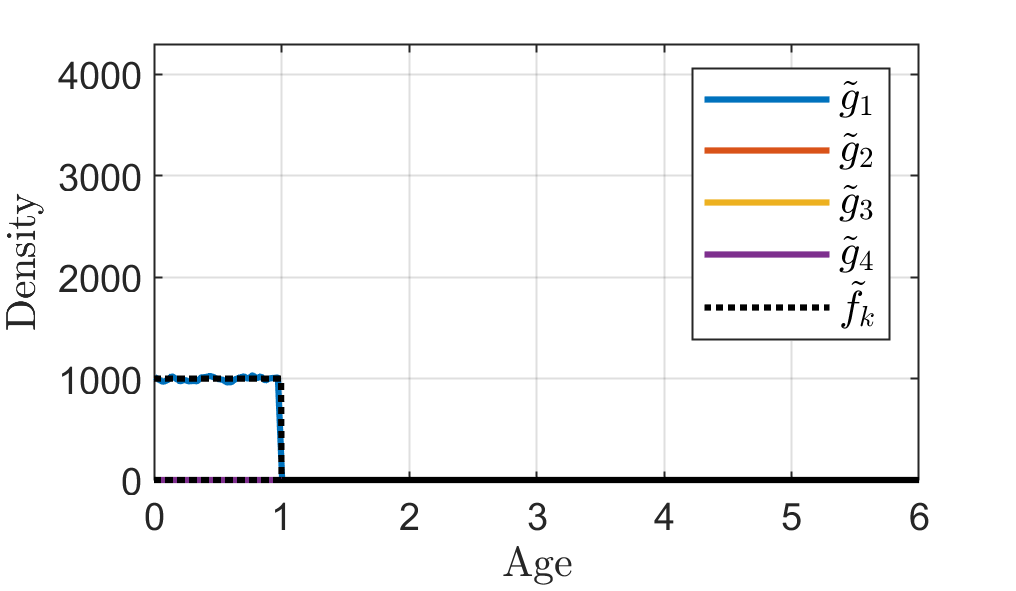}}\hfill
        \subfloat[]{\includegraphics[width=0.5\textwidth]{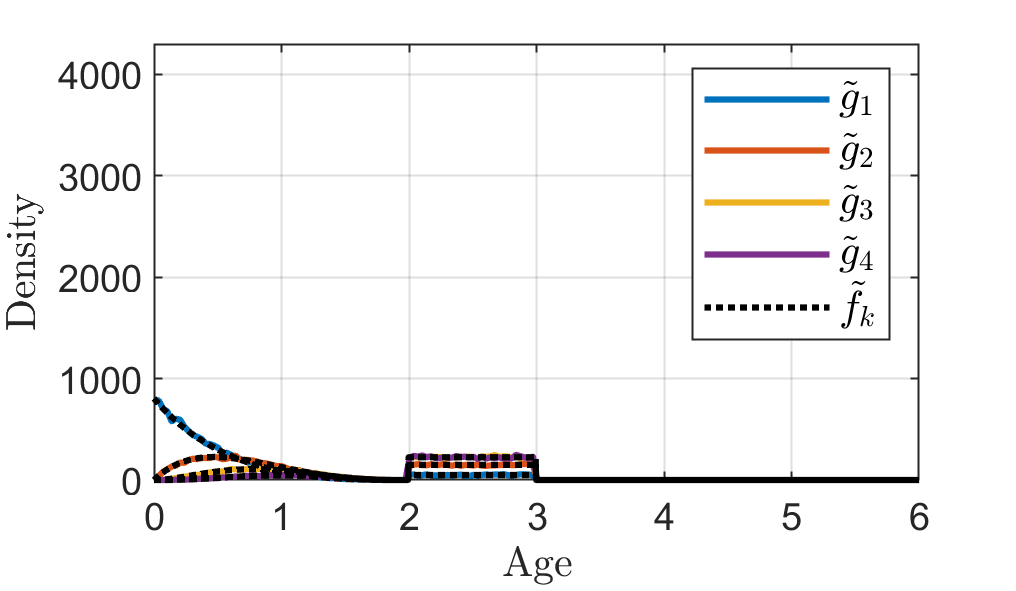}}
        \hspace{0em}
        \subfloat[]{\includegraphics[width=0.5\textwidth]{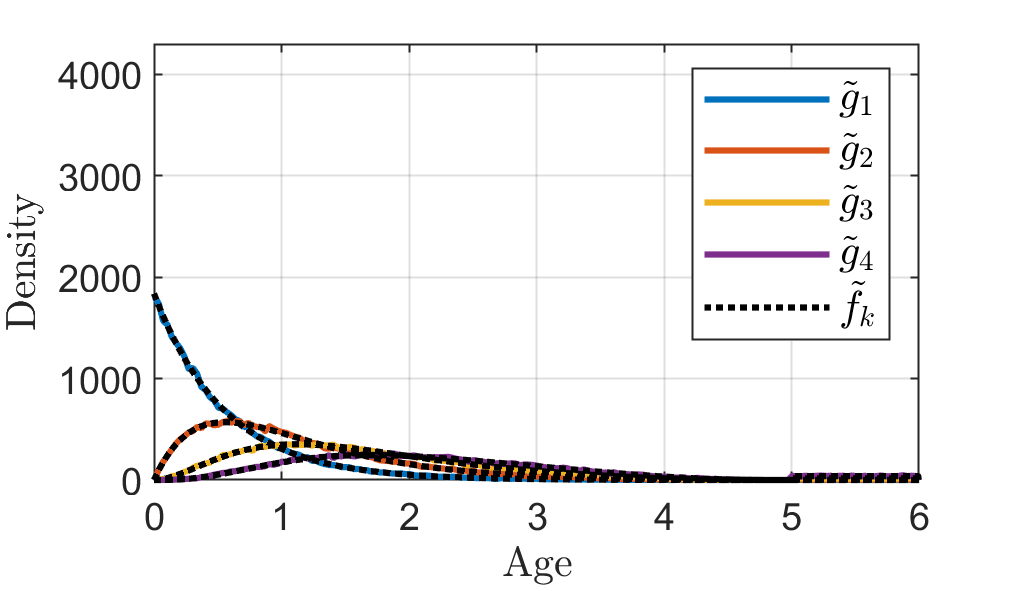}}\hfill
        \subfloat[]{\includegraphics[width=0.5\textwidth]{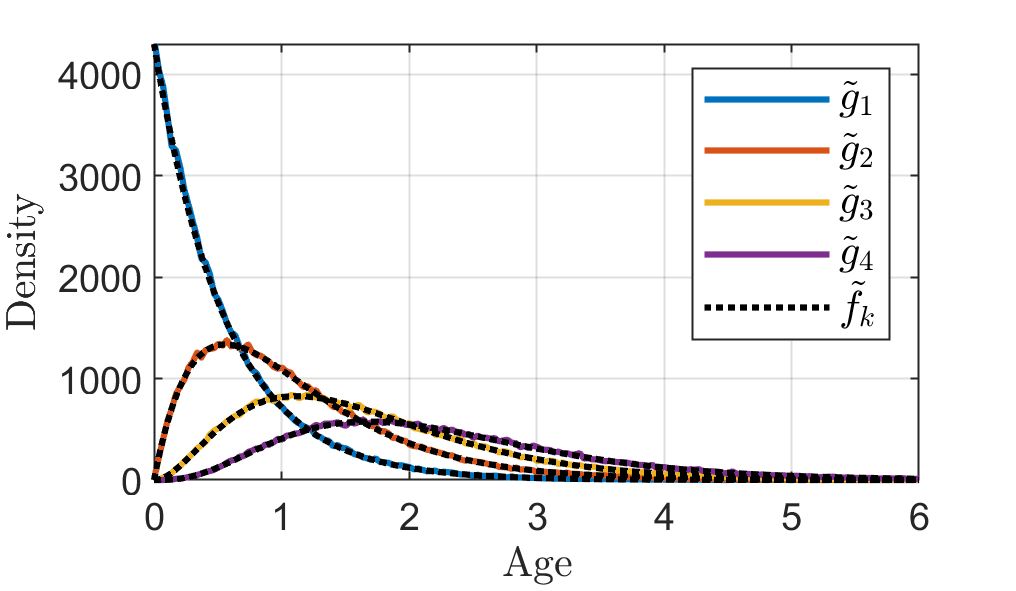}}
        \caption{
            Age densities in test problem 1, simulated with $\Delta\alpha = \frac{1}{300}$ and a time step of $\Delta t = \frac{1}{300}$ on the age domain $[0, 25]$. The displayed densities are plotted only on the domain $[0, 6]$ for ease of comparison. (a) initially, at time $t=0$, all cells are in pseudostage 1 with ages uniformly distributed in the interval $[0, 1]$. The black, dotted lines represent the corresponding numerical approximations to the densities of each stage. The remaining three panels (b), (c), and (d), display the propagation of the initial condition at (non-dimensional) times $t=2,4,8$, respectively. The coloured lines (described in the legend) correspond to the stochastic approximations to the densities of each of the four stages. Simulation results are averaged over 1000 repeats.
        }
        \label{fig:test1_densities}
    \end{figure}

    To be specific, consider a 4-stage aMSM with equal transition rates $\lambda = 1$. We take the initial condition
    \begin{equation}
        \tilde{f_k}(\alpha_i, 0) = \tilde{g_k}(\alpha_i, 0) = 
        \begin{dcases}
        1000 & \alpha_i \leq 1 \text{ and } k = 1, \\
        0 & \text{otherwise},
        \end{dcases}
    \end{equation}
    as illustrated in Fig.~\ref{fig:test1_densities}(a), corresponding to an initial population of $1000$ individuals in stage 1; we note the fact that this initial condition is incompatible with the boundary conditions~\eqref{eq:aMSM_bc}; nevertheless, our deterministic system clearly approximates the mean of the stochastic system, as we now demonstrate.
    
    \begin{figure}
        \centering
        \subfloat[]{\includegraphics[width=0.45\textwidth]{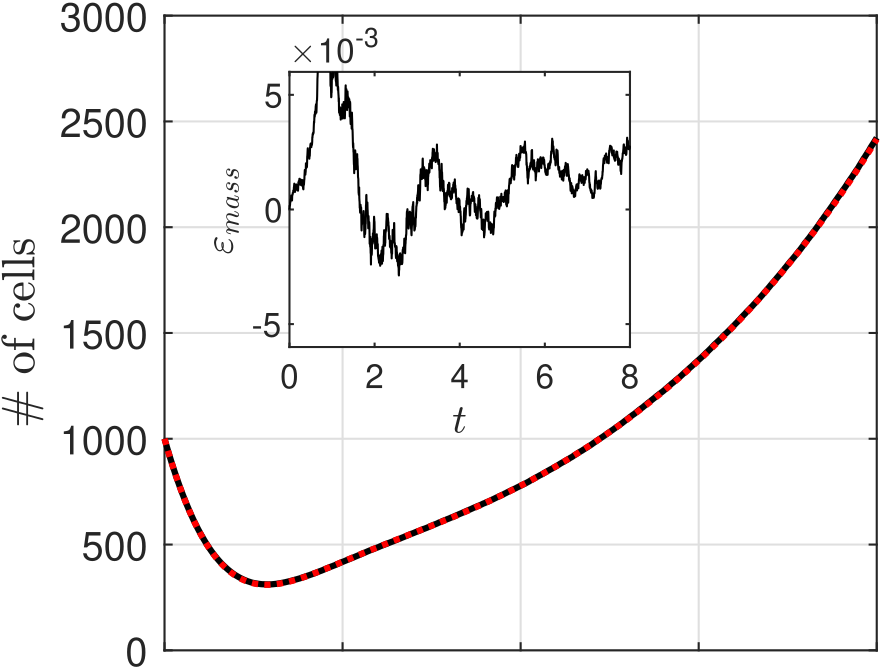}}\hfill
        \subfloat[]{\includegraphics[width=0.45\textwidth]{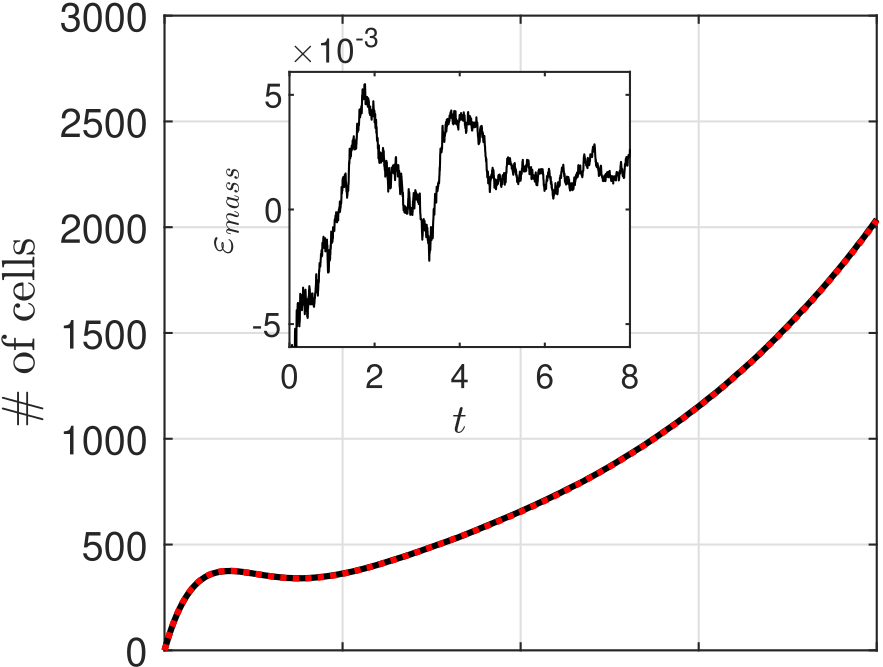}}
        \hspace{0em}
        \subfloat[]{\includegraphics[width=0.45\textwidth]{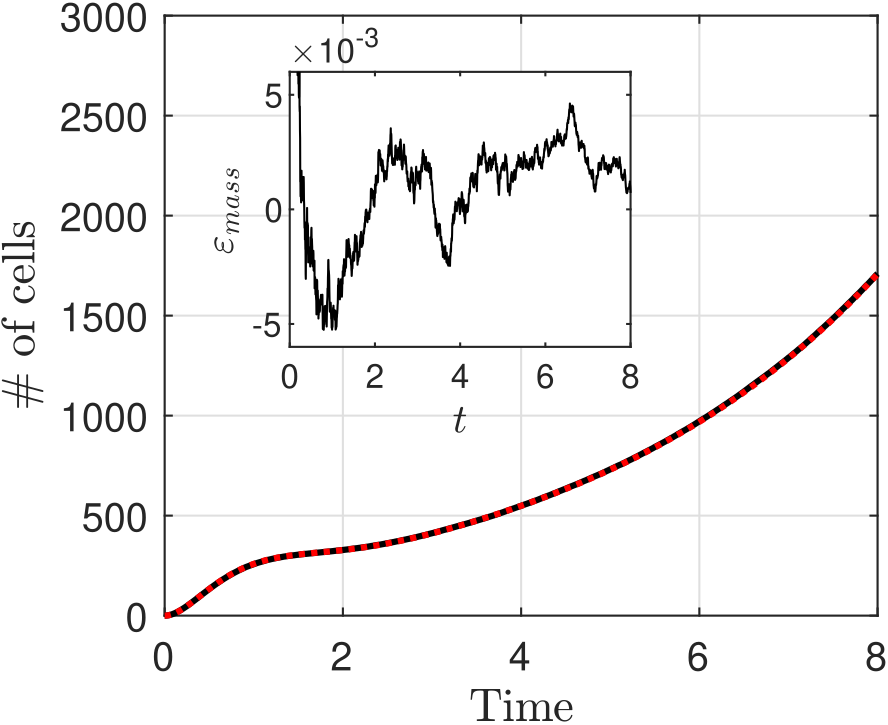}}\hfill
        \subfloat[]{\includegraphics[width=0.45\textwidth]{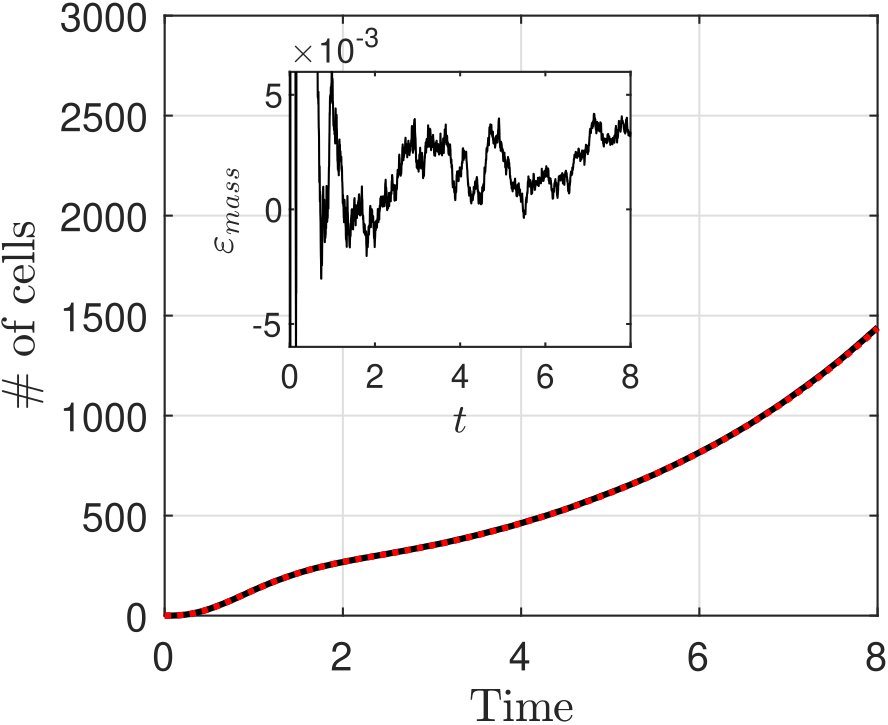}}
        \caption{Plots of the $1$-norm of the $\tilde f_k$ and $\tilde g_k$. Panel (a) shows the numerical approximation for the number of cells in pseudostage $1$ (black, dashed line) and the stochastic approximation (red, solid line). The inset plot shows the relative error over time; that is, $(\tilde g_1(t)- \tilde f_1(t))/\tilde g_1(t)$. The panels (b), (c), and (d) present the equivalent quantities for pseudostages 2, 3, and 4, respectively. Simulation results are averaged over 1000 repeats.}
        \label{fig:test1_mass_err}
    \end{figure}
    
    In the first example, shown in Fig.~\ref{fig:test1_densities}, we present a simple comparison of the stochastic and deterministic age densities in each stage at time points $t=0,2,4,8$. To the eye, these plots demonstrate excellent agreement between the stochastic and deterministic age densities. A more thorough investigation of this agreement is presented in Fig.~\ref{fig:test1_mass_err}. Here, we calculate and compare the integral over age of each curve $\tilde f_k(t)$ and $\tilde g_k(t)$ for $k=1,\hdots,4$. Since both functions are strictly positive, this can be interpreted as the number of cells present in each pseudostage. We also calculate the relative mass-error in each pseudostage, defining
    \begin{equation}
        \varepsilon_{mass}(t) := \frac{\int_0^\infty \tilde g_k(\sigma, t) - \tilde f_k(\sigma, t)\,\text d\sigma}{\int_0^\infty \tilde g_k(\sigma, t)\,\text d\sigma} \approx \frac{\sum_{i=0}^N (\tilde g_k(\alpha_i, t) - \tilde f_k(\alpha_i, t) )}{\sum_{i=0}^N \tilde g_k(\alpha_i,t)}.
    \end{equation}
    The panels in Fig.~\ref{fig:test1_mass_err} exhibit good agreement and no discernible systemic bias over large timescales (insets). 
    
\begin{figure}
        \centering
        \includegraphics[width=0.5\textwidth]{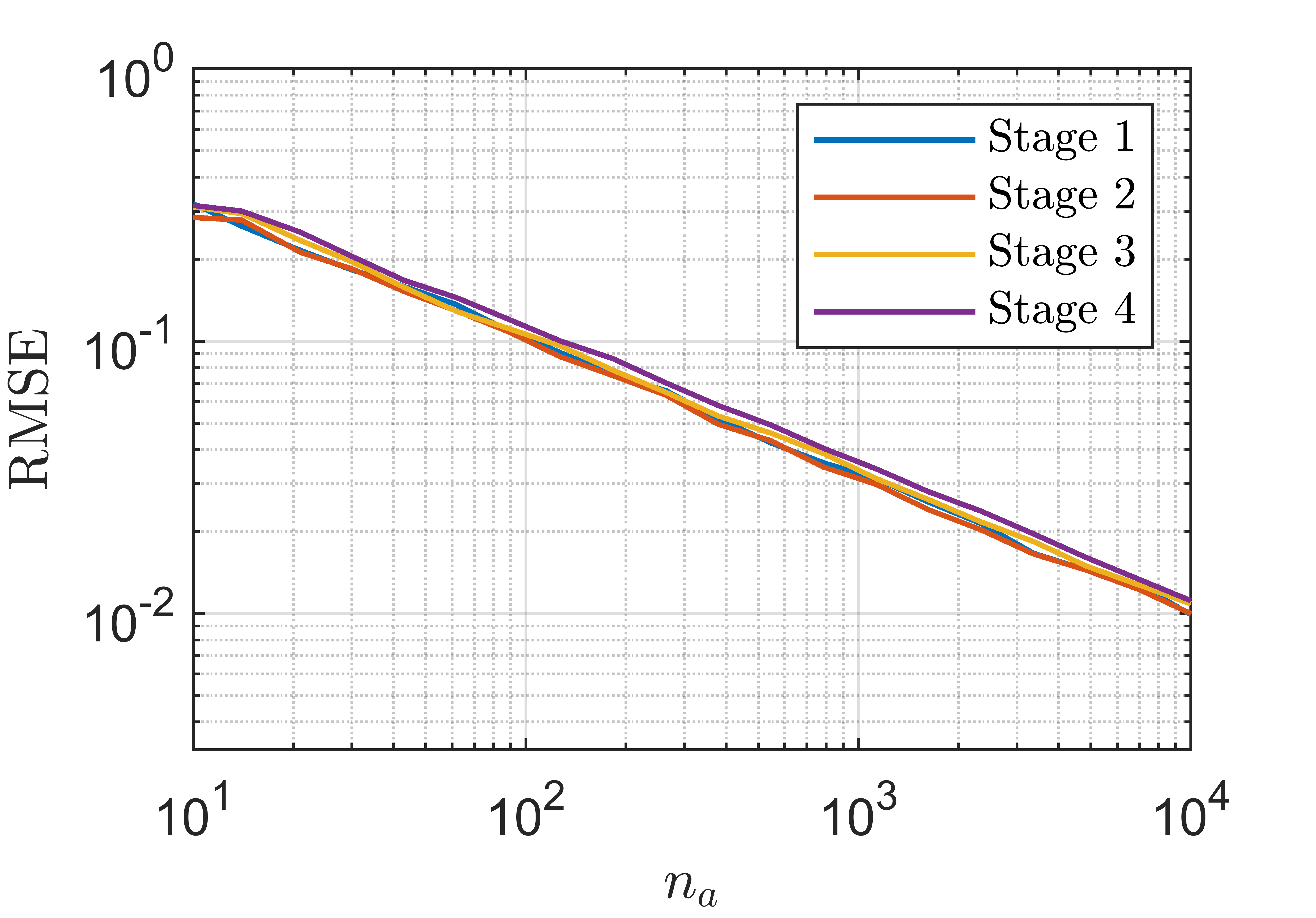}
        \caption{Plot of the RMSE between the $\tilde f_k$ and $\tilde g_k$ for a range of initial population sizes $n_a$ between $10^1$ and $10^4$ for each of the $K=4$ stages. MSE sampled at time $t=8$ and averaged over 1000 repeats.}
        \label{fig:rmse}
    \end{figure}

    To examine the effect of initial population size $n_a$ on model agreement, we consider the $\tilde f_k$ as predictors of $\tilde g_k$ in order to calculate the root mean squared error (RMSE) associated with the stochastic solution. We define the RMSE at time $t$ for stage $k$ via
    \[ \text{RMSE}_k(t)^2 = \frac{1}{\sum_{i=0}^N \tilde g_k (\alpha_i,t)} \sum_{i=0}^N (\tilde g_k(\alpha_i, t) - \tilde f_k (\alpha_i, t))^2.\]
    As demonstrated in a log-log plot (Fig.~\ref{fig:rmse}), we find strong evidence that the RMSE associated with our stochastic model decays with the reciprocal root of the initial population size. This is precisely the result one would expect from the central limit theorem for an unbiased predictor $\tilde f_k$, providing evidence of a lack of bias in our Monte Carlo simulation.
    
\begin{figure}
        \centering
        \subfloat[][]{\includegraphics[width=0.5\textwidth]{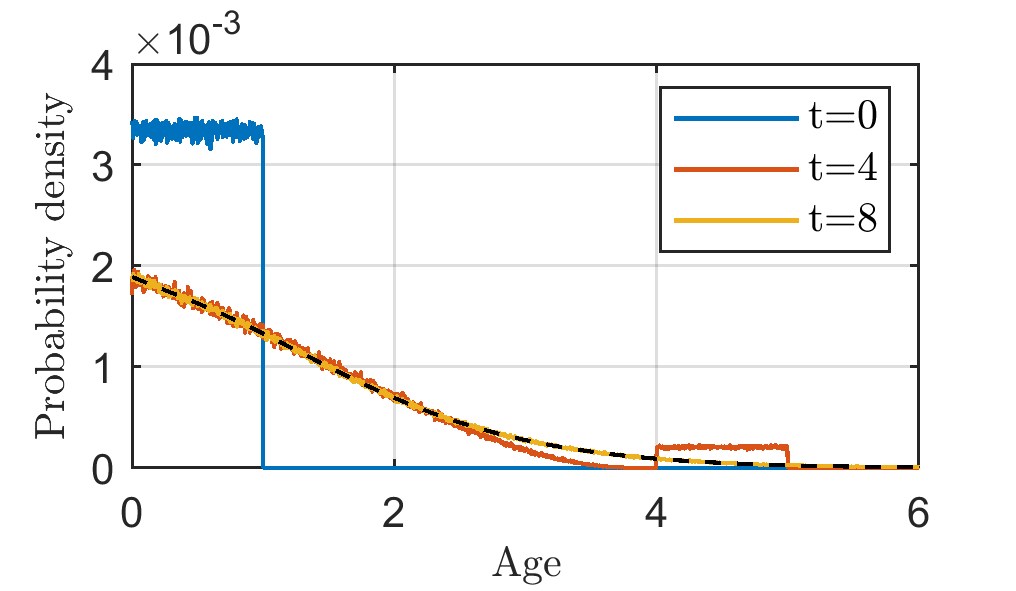}}
        \subfloat[][]{\includegraphics[width=0.5\textwidth]{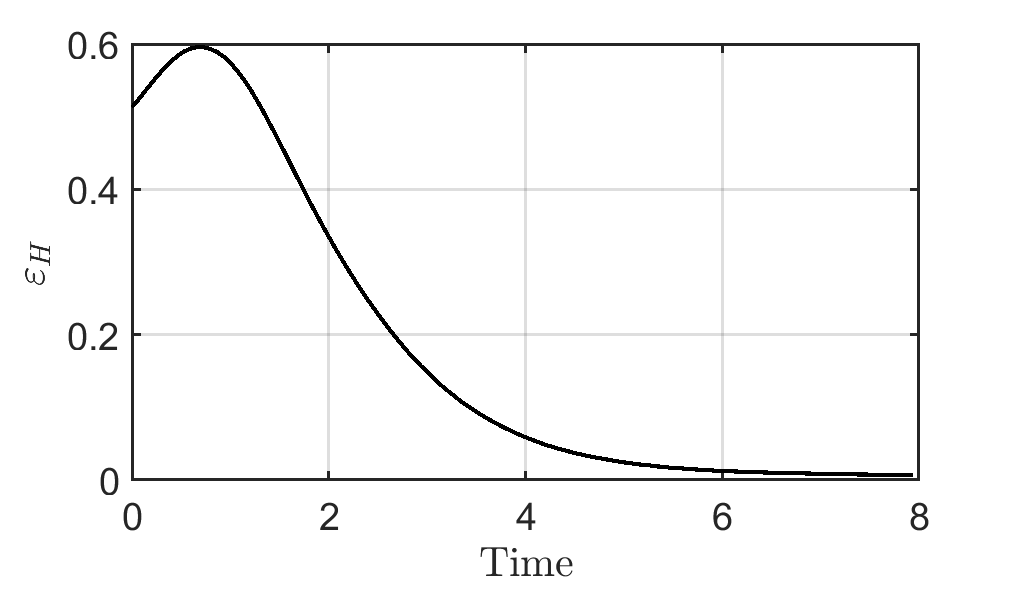}}
        \caption{(a) Plot of the combined age distribution for three time points $t=0, 6, 12$ as approximated via a stochastic simulation of~\eqref{eq:aMSM} versus the deterministic asymptotic distribution~\eqref{eq:exact_combined_dist} (black, dashed). (b) Evolution of the histogram distance error between the stochastic approximation of the combined age distribution and the deterministic asymptotic distribution over time. Simulation results are averaged over 1000 repeats.}
        \label{fig:combined_dist}
    \end{figure}

    Finally, we examine the convergence of the approximate combined age distribution~\eqref{eq:approx_combined_dist} to its theoretical limit~\eqref{eq:exact_combined_dist}. To this end, we employ the histogram distance error (HDE) metric \cite{cao_accuracy_2006}, here defined to be
    \begin{equation}
        \varepsilon_H(t) := \frac{1}{2}\sum_{i=0}^N \left|\frac{\tilde\zeta(\alpha_i, t)}{\sum_{j=0}^N \hat \zeta(\alpha_j, t)} - \frac{\hat \zeta^*(\alpha_i)}{\sum_{j=0}^N \hat \zeta^*(\alpha_j)}\right|.
    \end{equation}
    The HDE takes values in $[0, 1]$, where $0$ corresponds to the two distributions being equal almost everywhere, and a value of $1$ indicates that the supports of the two distributions are disjoint. We plot the HDE in Fig.~\ref{fig:combined_dist}, observing a rapid convergence of the combined age distribution to its theoretical limit.
    
\subsection{Age and space structure}
We now extend the results of the previous sections to a spatial context. Incorporating spatial effects into the aMSM could be accomplished in several ways; however, in this test problem we consider a one-dimensional `mesoscopic' representation, where the spatial location of individuals is discretised into fixed-sized compartments. This approach has the advantage that it can be simulated quickly, and is easily implemented via the Gillespie algorithm through the introduction of compartments to the well-mixed version of the aMSM. This approach is also amenable to hybridisation via approaches such as the psuedo-compartment method \cite{yates_pseudo-compartment_2015} or the method of Spill \textit{et al.} \cite{spill_hybrid_2015}.

The (mesoscopic) spatial $k$-stage aMSM is a continuous-time agent-based method on a finite domain, which we will take without loss of generality to be $\Omega_C = \mathbb [0, 1]$, partitioned into $N_c$ compartments of uniform width $h$, with central points $x_j = (2j-1)h/2$ for $j=1,...,N_c$. Individuals within each compartment have an associated stage, $k$. As in the well-mixed model (i.e., the model without spatial structure), stage $k$ individuals can transition to stage $k+1$ at a rate $\lambda$. If an individual is in stage $K$, then at rate $\lambda$ it undergoes a cytokinesis event, where it transitions back to stage $1$ and produces another individual in stage $1$ in the same compartment. Diffusion is simulated by permitting individuals to move to adjacent compartments at a rate $D/h^2$, where $D$ is the Fickian diffusion coefficient. Various boundary behaviours can be realised by suitably modifying transition rates within boundary compartments, such as adsorbing boundaries \cite{erban_reactive_2007}. The simplest, however, is the zero-flux boundary, where individuals attempting to jump out of the domain are simply reflected back into their originating compartment, which is what we shall consider here.

A full derivation, from first principles, of the mean age and space structure of the spatial aMSM is unnecessary since (pseudo)stage transition events are entirely decoupled from diffusive jump events between compartments. The approximate age and space density of the spatial aMSM takes the form
\begin{align}
    \frac{\partial q_k}{\partial t} + \frac{\partial q_k}{\partial a} - D \frac{\partial^2 q_k}{\partial x^2} &= \begin{dcases}
    -\lambda q_k & k = 1, \\
    -\lambda q_k + \lambda q_{k-1} & k = 2,\hdots,K,
    \end{dcases} \label{eq:asMSM_pde} \\
    q_k(0, x, t) &= \begin{dcases}
    2\lambda \int_0^\infty q_k(\sigma, x, t)\,\text d\sigma & k = 1, \\
    0 & k = 2,\hdots,K,
    \end{dcases} \label{eq:asMSM_abc}\\
    \frac{\partial q_k}{\partial x}(a, 0, t) &= \frac{\partial q_k}{\partial x}(a, 1, t) = 0, \label{eq:asMSM_sbc}
\end{align}
where $q_k(a,x,t)$ gives the density of individuals with age $a \in [a + \text da)$ in $x \in [x + \text dx)$ at time $t$. From this, we can define the combined age density
\begin{equation}
    \xi (a,x,t) = \sum_{i=1}^K q_k(a,x,t).
\end{equation}
In particular, when the initial conditions $q_k(a,x,0)$ satisfy~\eqref{eq:asMSM_pde} and~\eqref{eq:asMSM_abc}, then $\xi$ is a solution to the following initial boundary value problem,
\begin{align}
    \begin{split}
        \frac{\partial \xi}{\partial t} + \frac{\partial \xi}{\partial a} &= D \frac{\partial^2 \xi}{\partial x^2} - \lambda\xi, \\
        \xi(0,x,t) &= 2 \lambda \int_0^\infty \xi(\sigma,x,t)\,\text d\sigma, \\
        \frac{\partial}{\partial x} \xi(a,0,t) &= \frac{\partial}{\partial x} \xi(a,1,t) = 0.
    \end{split}\label{eq:smvfe}
\end{align}
Note that this problem can be viewed as a spatial extension of the MVFE \cite{langlais_large_1988}. Since \eqref{eq:smvfe} can be written as a linear combination of differential operators, we can calculate analytical solutions to \eqref{eq:smvfe} by taking the product of solutions to the canonical diffusion equation with zero-flux boundaries and the MVFE. Specifically, it is easily verified that if $\rho(a,t)$ is a solution to \eqref{eq:mvfe}-\eqref{eq:mvfebc}, and $\psi(x,t)$ is a solution to
\begin{align*}
    \frac{\partial \psi}{\partial x} &= D \frac{\partial^2 \psi}{\partial x^2}, \\
    \frac{\partial}{\partial x}(0,t) &= \frac{\partial}{\partial x}(1,t) = 0,
\end{align*}
then their product $\xi(a,x,t) = \rho(a,t)\psi(x,t)$ is a solution to \eqref{eq:smvfe}.

\begin{figure}
    \centering
    \subfloat[]{\includegraphics[width=.45\textwidth]{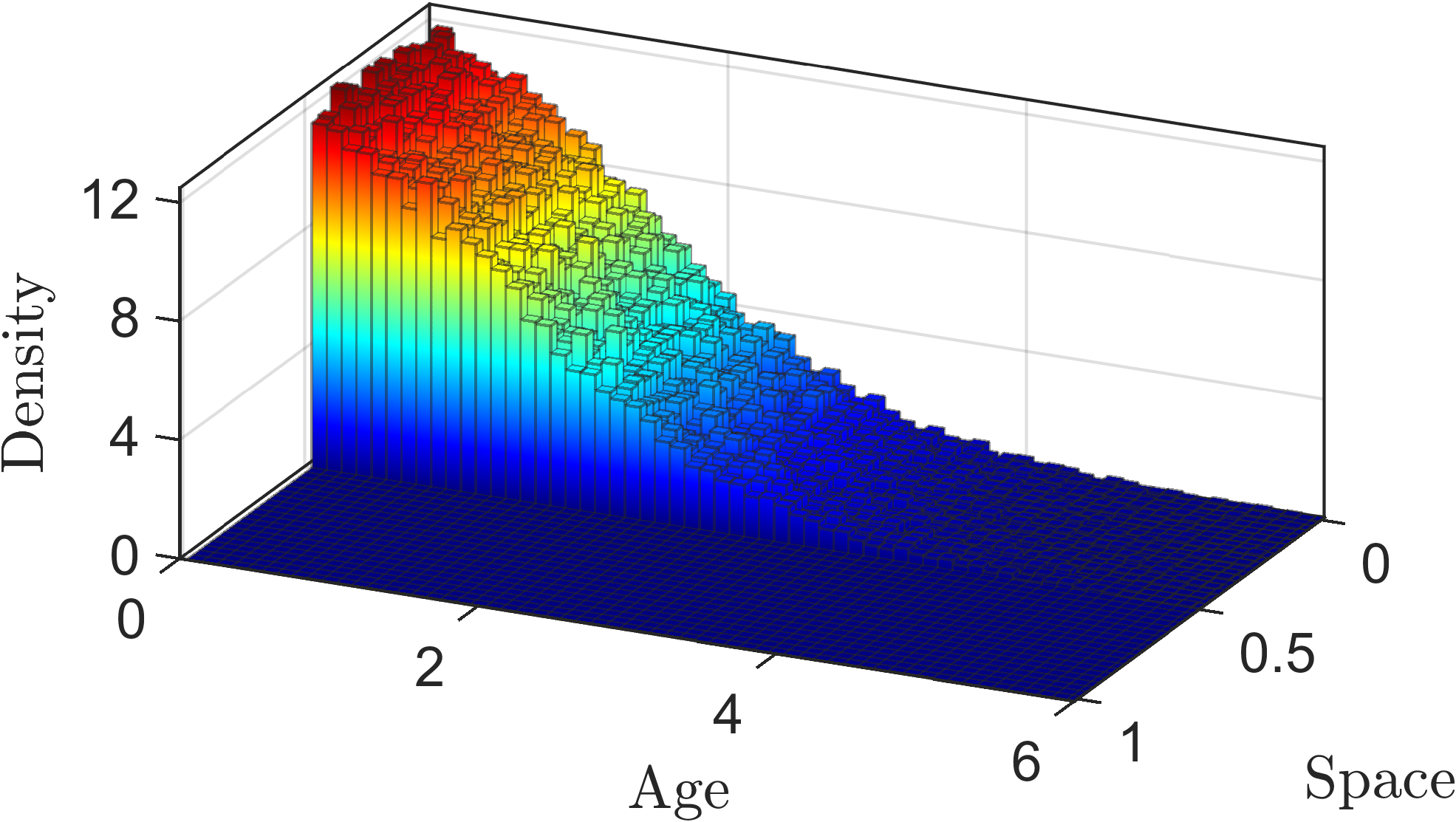}}\hfill
    \subfloat[]{\includegraphics[width=.45\textwidth]{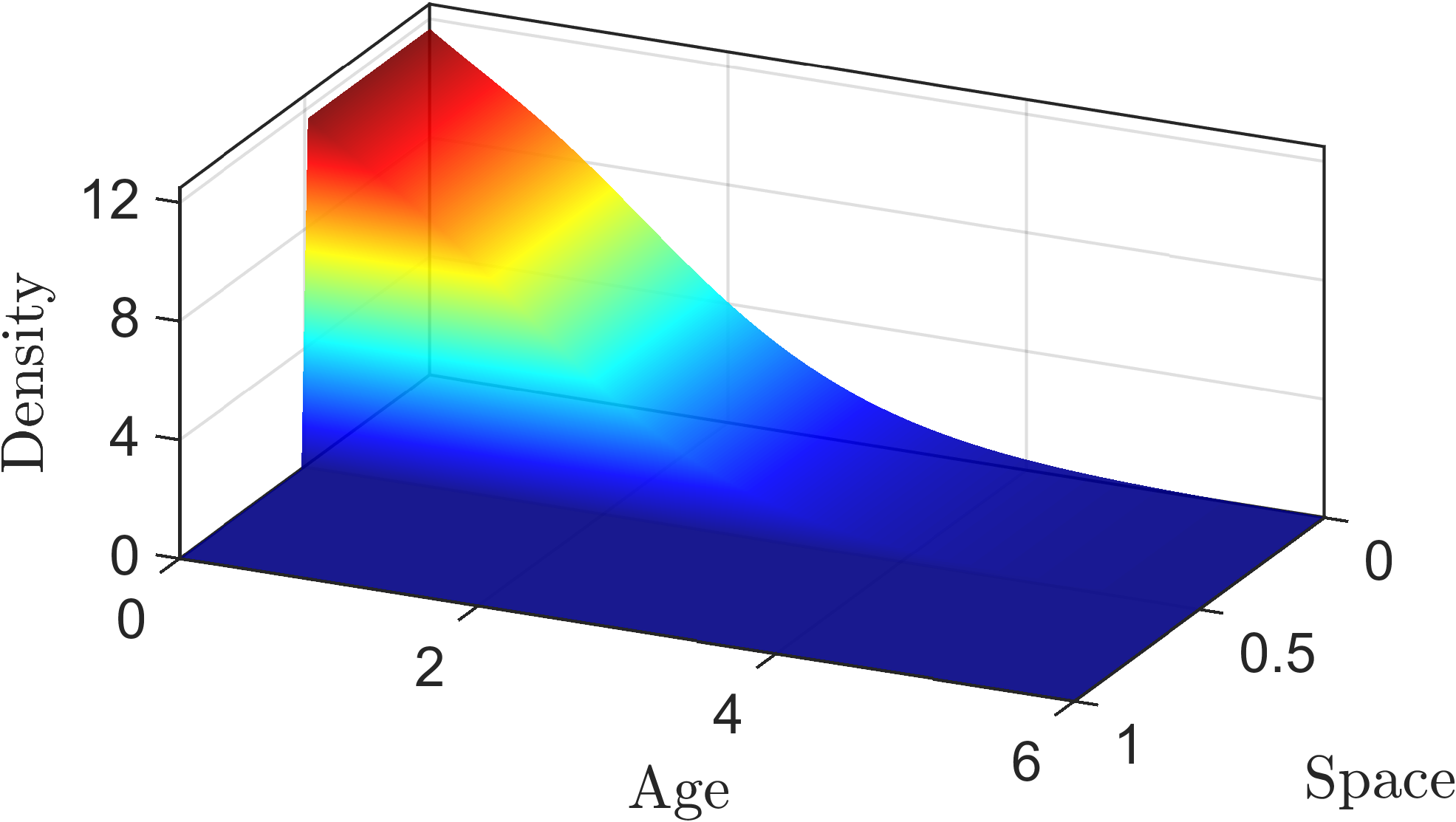}}\vfill
    \subfloat[]{\includegraphics[width=.45\textwidth]{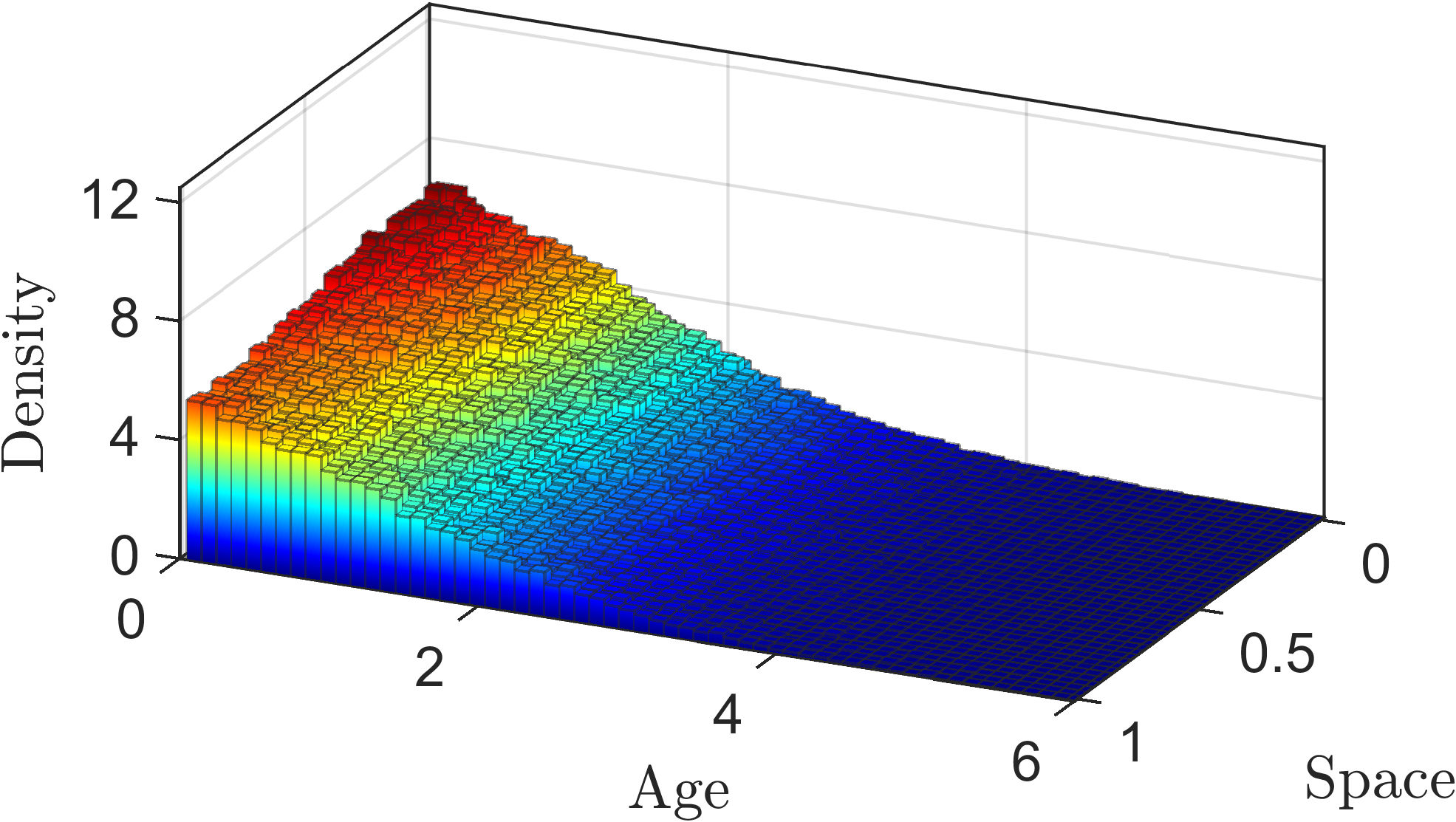}}\hfill
    \subfloat[]{\includegraphics[width=.45\textwidth]{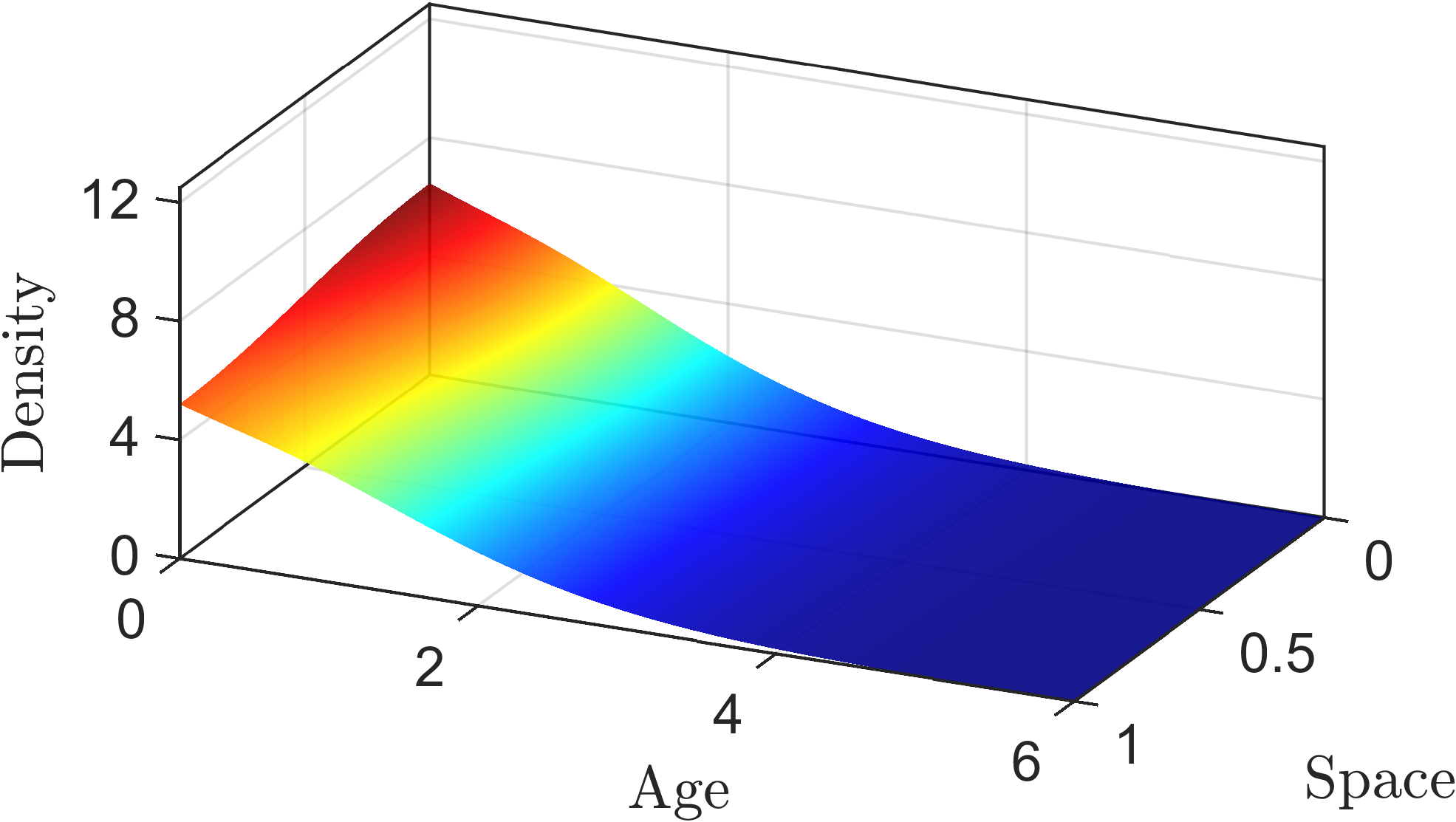}}
    \caption{A presentation of the age and space distribution of the spatial aMSM with transition rate $\lambda = 2$ and diffusion coefficient $D = \frac{1}{20}$. Top row: Initial condition specified by equation \eqref{eq:2d_ic}. Bottom row: propagation of the initial condition over 5 units of (non-dimensional) time. Left column: The plotted histograms represent the density of cells in each compartment of the domain and in each age interval, with compartment width $h=\frac{1}{40}$ and age step $\Delta \alpha = \frac{1}{40}$. Right column: Surface plot of the analytical solution for the age and space density, with $\Delta x = \frac{1}{200}$, $\Delta \alpha = \frac{1}{40}$, and time step $\Delta t = \frac{1}{100}$. Simulation results are averaged over 100 repeats.}
    \label{fig:age_space_densities}
\end{figure}

Denote by $\tilde q_k$ the age density of the $k^\text{th}$ stage of the spatial aMSM as approximated via stochastic simulation, and define $\tilde \xi = \sum_{k=1}^K \tilde q_k$. We take the initial condition
\begin{equation}
    \tilde q_k(\alpha_i,x_j,0) = \begin{dcases}
    100 \lim_{t\rightarrow\infty}\{P_k(t)\} \hat f^*_k(\alpha_i) & x_j \leq 0.5, \\
    0 & x_j > 0.5,
    \end{dcases}\label{eq:2d_ic}
\end{equation}

recalling the definitions of $P_k(t)$ and $\hat f^*_k$ from~\eqref{eq:prop_definition} and~\eqref{eq:stage_density_limit_def}, respectively. This initial condition can be interpreted as each (non-empty) compartment containing 100 cells which have already reached a persistent distribution of ages. This type of initial condition is a common experimental setup; for example, in proliferation assays, where the spread of an established cellular population is observed, and in so-called scratch assays, where a `wound' is made in a cell monolayer, and the migration of cells into the empty region is observed (see \cite{torisawa_proliferation_2004,rodriguez_wound-healing_2005,huyck_xtt_2012}, for example).

\begin{figure}
    \centering
    \includegraphics{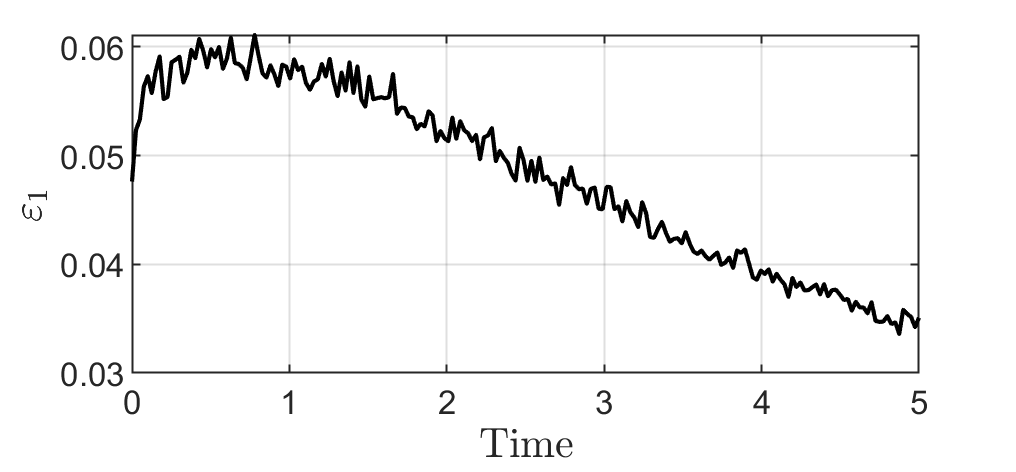}
    \caption{Distance in the 1-norm between the combined age density of the spatial aMSM as approximated via stochastic simulation and the analytical density. Simulation results are averaged over 100 repeats.}
    \label{fig:1norm_2d}
\end{figure}

Figure~\ref{fig:age_space_densities} illustrates the evolution of the spatial aMSM alongside its continuum approximation. For the purposes of evaluating the degree to which $\xi$ approximates $\tilde\xi$, the HDE is no longer a suitable metric. Since neither $\xi$ nor $\tilde\xi$ are normalised, to compare them as probability distributions over age and space could be misleading, since we expect the population to grow exponentially over time. Instead, we consider the approximate difference in the 1-norm between the two surfaces,

\begin{equation}
    \varepsilon_1(t) = \sum_{i=0}^N \sum_{j=1}^{N_c} | \tilde\xi(\alpha_i, x_j, t) - \xi(\alpha_i, x_j, t) |\Delta \alpha \Delta x,
\end{equation}

which we present in Fig \ref{fig:1norm_2d}. Our results demonstrate good agreement between the stochastic spatial aMSM and its counterpart; indeed, this agreement tends to improve over time. We observe a worsening of the error toward the beginning of the simulation run due to numerical discrepancies between the two approximations. In particular, the initial step function in space is instantaneously smoothed by the diffusion operator in the partial differential equations after $t=0$, whereas individuals in the stochastic method have finite velocity in space, and therefore we find more density near $x=1$ in the deterministic versus the stochastic method. This discrepancy, however, improves over time.

\section{\centering \MakeUppercase{Discussion}}
    The present work describes the complete age and time evolution of the multi-stage model~\cite{yates_multi-stage_2017}. The key mathematical results of this work, namely equations~\eqref{aMSM_MARG} and \eqref{aMSMBC_MARG}, describe the full age-structured dynamics of the multi-stage model and can be simplified to extract key statistics such as the mean age density~\eqref{eq:aMSM_pde}-\eqref{eq:aMSM_bc}. The power of the multi-stage model is its ability to model non-exponential cell cycle time distributions via the Gillespie algorithm~\cite{gillespie_exact_1977} by breaking down the cell cycle into a series of independent, exponentially distributed, stages. We have shown how the limiting age distribution of the multi-stage model is related to the CCTD~\eqref{eq:exact_combined_dist}, and qualitatively evaluated the accuracy of the continuum description through two test cases that directly compare the mean of stochastic simulations of the age-structured multi-stage model with numerical approximations of the mean age density, in both a non-spatial and spatial context. We find, in both cases, that our continuum theory is an excellent predictor for the underlying stochastic process. 
    
    The are several natural extensions to our theory. The first is the incorporation of population-level effects including volume exclusion and cell-to-cell adhesion. These processes form a critical element for understanding and constructing effective models for tumour growth and metastatic potential in breast carcinomas \cite{vultur_cell--cell_2004,simpson_migration_2010} and melanomas \cite{mcgary_cellular_2002}, as well as for understanding chemotherapeutic resistance \cite{garrido_hsp27_1997}, for example. A further extension is for the incorporation of a more diverse reaction set into the aMSM. An individual cell does not exist independently of its neighbours, and may transmit chemical signals in response to local environmental conditions \cite{ratajczak_membrane-derived_2006}. For example, \textit{Psuedomonas aeruginosa} are known to transmit cell-to-cell signals in the formation of biofilms, resulting in populations that are morphologically distinct from other bacterial colonies which do not interact in this manner \cite{davies_involvement_1998}. Further, while the aMSM can resolve the age-structure of individual phases of the cell cycle, there are other important variables through which one can study cellular populations, such as via size or concentration metrics \cite{xia_kinetic_2021}. Whether the conceptual framework of the aMSM as a series of memoryless stages is equipped to incorporate size warrants further inquiry. Finally, the inclusion of multiple species into the model would be of great benefit to its utility. One natural application of this would be for modelling stem cell populations undergoing differentiation in the development of stem cell therapies, for which age structure is an important element \cite{zakrzewski_stem_2019}. Further, a deterministic multi-species age-structured approach has been used for modelling phenotypic plasticity in tumour growth \cite{cassidy_role_2021}, representing a potential extension and subsequent application for our single-species equivalence framework.

\section*{\centering \MakeUppercase{Acknowledgements}}
This research made use of the Balena High Performance Computing (HPC) Service at the University of Bath. Joshua C. Kynaston is supported by a scholarship from the EPSRC Centre for Doctoral Training in Statistical Applied Mathematics at Bath (SAMBa), under the project EP/L015684/1.

\appendix

\section{Derivation of the marginal densities} \label{sec:supp_1}
In this appendix we demonstrate the techniques for deriving the marginal density equations~\eqref{aMSM_MARG}~and~\eqref{aMSMBC_MARG}. Consider the following master equation
\begin{align}
\begin{split}
    \frac{\partial \rho_{\bm n}}{\partial t} + \sum_{k=1}^{K+1} \sum_{i=1}^{n_k} \frac{\partial \rho_{\bm n}}{\partial (a_k)_i} = &\sum_{k=1}^{K} \lambda_k \left ( \frac{n_k + 1}{n_{k+1}} \right) \sum_{i=1}^{n_{k+1}} \rho_{\mathcal F_k \bm n}\left\{\hdots\,; \bm a_{n_k}, (a_{k+1})_i \,; \bm a^{(-i)}_{n_{k+1}}\,; \hdots \,; t \right\} \\
    &+ 2\lambda_1 \left(\frac{n_{K+1} + 1}{n_1 n_2}\right) \sum_{i=1}^{n_1} \sum_{j=1}^{n_2} \rho_{\mathcal G \bm n} \left\{\bm a_{n_1}^{(-i)}\,; \bm a_{n_2}^{(-j)}\,;\, \hdots\,; \bm a_{n_{K+1}}, \omega_{i,j}\, ; t \right\} \\
    &-\left(2 \lambda_1 n_{K+1} + \sum_{k=1}^{K} \lambda_k n_k \right) \rho_{\bm n}, \\
\end{split} \label{seq:master_eqn}
\end{align}
with boundary conditions
\begin{align}
\begin{split}
    \rho_{\bm n}\left\{\hdots\,; \bm a_{n_k - 1}, 0\,;\hdots;t\right\} = \begin{dcases}
    \lambda_K \left( \frac{n_K + 1}{n_{K+1}} \right) \int_0^\infty \rho_{\mathcal F_K \bm n} \left\{\hdots\,;\bm a_{n_K}, \sigma\,;\hdots\,; t \right\}\,\text d\sigma & k = K+1, \\
    0 & k = 1,\hdots, K.
        \end{dcases} 
\end{split}\label{seq:master_eqn_bc}
\end{align}
Define the following multi-integral operator
\begin{equation*}
    \sI^{\bm m} := \int_{0}^\infty \hdots \int_0^\infty \text d \bm a'_{n_1 - m_1} \hdots \text d \bm a'_{n_{K+1} - m_{K+1}},
\end{equation*}
recalling that
\[ \rho_{\bm n}^{\bm m} = \sI^{\bm m} (\rho_{\bm n}). \]
We wish to determine the result of applying the operator $\sI^{\bm m}$ to the master equation~\eqref{seq:master_eqn} and boundary conditions~\eqref{seq:master_eqn_bc}. Beginning with the boundary conditions, observe that
\[ \sI^{\bm m} ( \rho_{\bm n} \{\hdots;\, a_{n_k - 1}, \sigma;\hdots;t\} ) = 0\quad \text{for } k = 1,\hdots,K,\]
leaving the case where $k = K+1$, for which
\begin{align*}
     \sI^{\bm m} ( \rho_{\bm n} \{\hdots;\, a_{n_{K+1} - 1}, \sigma;t\} ) &= \lambda_K \left(\frac{n_K + 1}{n_{K+1}} \right) \int_0^\infty \sI^{\bm m} \left[ \rho_{\sF_K \bm n}\{\hdots;a_{n_K},\sigma;\hdots;t \} \right] \\
    &= \lambda_K \left( \frac{n_K + 1}{n_{K+1}} \right) \int_0^\infty \rho_{\sF_K \bm n}^{\bm m}\{\hdots;a_{m_K}, \sigma;\hdots;t \}\,\text d\sigma \\
    &= \lambda_K \left( \frac{n_K + 1}{n_{K+1}} \right)\rho_{\sF_K \bm n}^{\sH_K^- \bm m}.
\end{align*}
Next we deal with the partial derivatives in age. Assign the labels (\ref{seq:master_eqn}.1) and (\ref{seq:master_eqn}.2) to the two terms on the left hand side of~\eqref{seq:master_eqn} respectively; likewise, assign (\ref{seq:master_eqn}.3), (\ref{seq:master_eqn}.4), and (\ref{seq:master_eqn}.5) to the three terms on the right hand side. We begin with (\ref{seq:master_eqn}.1), for which it trivially holds that
\[(\text{\ref{seq:master_eqn}}.1) = \sI_{\bm n}^{\bm m} \left( \frac{\partial \rho_{\bm n}}{\partial t} \right) = \frac{\partial \rho_{\bm n}^{\bm m}}{\partial t}. \]
The integration of (\ref{seq:master_eqn}.2) more care. Specifically, we split the inner sum into two parts, first summing over variables which are not integrated out by $\sI^{\bm m}$, and then summing over those that are,
\begin{align*}
    \sI^{\bm m}\left( \sum_{k=1}^{K+1} \sum_{i=1}^{n_k} \frac{\partial \rho_{\bm n}}{\partial (a_k)_i} \right) &= \sI^{\bm m}\left( \sum_{k=1}^{K+1}\left[ \sum_{i=1}^{m_k} \frac{\partial \rho_{\bm n}}{\partial (a_k)_i} + \sum_{j=m_k + 1}^{n_k} \frac{\partial \rho_{\bm n}}{\partial (a_k)_j} \right]\right) \\
    &= \sum_{k=1}^{K+1} \sum_{i=1}^{m_k} \sI^{\bm m} \left( \frac{\partial \rho_{\bm n}}{\partial (a_k)_i} \right) + \sum_{k=1}^{K+1} \sum_{j=m_k + 1}^{n_k} \sI^{\bm m} \left( \frac{\partial \rho_{\bm n}}{\partial (a_k)_j} \right) \\ 
    &= \sum_{k=1}^{K+1} \sum_{i=1}^{m_k} \frac{\partial \rho_{\bm n}^{\bm m}}{\partial (a_k)_i} + \sum_{k=1}^{K+1} \sum_{j=m_k + 1}^{n_k} \sI^{\bm m} \left( \frac{\partial \rho_{\bm n}}{\partial (a_k)_j} \right),
\end{align*}
where the second step follows from linearity, and the final step follows from the fundamental theorem of calculus. It remains then to calculate the integrals of the partial age derivatives of $\rho_{\bm n}$. We have, for $j = m_k+1, \hdots, n_k$ and $k = 1,\hdots,K+1$,
\begin{align*}
    \sI^{\bm m}\left( \frac{\partial \rho_{\bm n}}{\partial (a_k)_j}\right) &= \int_0^\infty \frac{\partial}{\partial (a_k)_j} \sI^{\sH_{k}^{+} \bm m}\left(\rho_{\bm n}\right)\,\text d(a_k)_j \\
    &= \int_0^\infty \frac{\partial}{\partial (a_k)_j} \rho^{\sH_k^+ \bm m}_{\bm n}\,\text d (a_k)_j \\
    &= - \rho_{\bm n}^{\sH_k^+ \bm m}\{ \hdots; a_{m_k}, 0; \hdots ; t \}
\end{align*}
where the first step arises from multiple applications of Leibniz' rule for integration, and the third step follows from an application of the fundamental theorem of calculus, making note that no cells can have infinite age and therefore
\[ \lim_{y\rightarrow\infty}\rho_{\bm n}^{\sH_k^+ \bm m}\{\hdots;a_{m_k}, y; \hdots ;t\} = 0. \]
Finally, note that
\[ - \rho_{\bm n}^{\sH^+_k \bm m}\{ \hdots; a_{m_k}, 0; \hdots ; t \} = 
\begin{dcases}
\lambda_K \left(\frac{n_K + 1}{n_{K+1}}\right) \rho_{\sF_K \bm n}^{\bm m} & k = K+1, \\
0 & k = 1,\hdots,K,
\end{dcases}
\]
which follows from the boundary conditions. Therefore,
\[ (\ref{seq:master_eqn}.2) = \sI^{\bm m}\left( \sum_{k=1}^{K+1} \sum_{i=1}^{n_k} \frac{\partial \rho_{\bm n}}{\partial (a_k)_i} \right) = \sum_{k=1}^{K+1}\sum_{i=1}^{m_k} \frac{\partial \rho_{\bm n}^{\bm m}}{\partial(a_k)_i} - \lambda_K (n_{K+1} - m_{K+1}) \left(\frac{n_K + 1}{n_{K+1}}\right) \rho_{\sF_K \bm n}^{\bm m}.\]
Applying similar splitting procedures to (\ref{seq:master_eqn}.3) - (\ref{seq:master_eqn}.5) yields the full marginal master equation \eqref{aMSM_MARG}.
\section{Long-term exponential growth}\label{sec:ode_limit}
Consider the system of linear ordinary differential equations
\begin{align}
\begin{split}
    \frac{\, \text dM_1}{\, \text d t} &= -\lambda M_1 + 2\lambda M_K, \\
    \frac{\, \text dM_k}{\, \text d t} &= -\lambda M_k + \lambda M_{k-1}, \quad \text{for $k = 2,...,K$},
\end{split}\label{ode_system}
\end{align}
with some initial condition $\bM(0) = (M_1(0), ..., M_K(0))$.
Setting
\[ A := 
\bpm{-\lambda &0&\cdots&2 \lambda \\\lambda &-\lambda &\ddots&\vdots\\&\ddots&\ddots&0\\0&& \lambda &-\lambda } \quad \text{and} \quad \bM := \bpm{M_1 \\ \vdots \\ M_K}\,,
\]
the system of differential equations~\eqref{ode_system} can be written in first order form
\[ \dot \bM(t) = A \bM(t)\,. \]
The matrix $A$ is Metzler, as all off-diagonal entries are nonnegative. It is a standard exercise to verify that the spectral abscissa of $A$ equals
\[ r := \lambda\big(2^{\frac{1}{K}} - 1 \big)\,.\]
As the matrix $A$ is additionally irreducible, it follows from, for example~\cite[Lemma 3.1]{bill_stability_2016} that provided at least one of the $M_k(0)$ is positive, then
\[ M_k(t) >0 \quad \text{for all $t > 0$, for all $k = 1, \dots, K$}.\]
Moreover, an application of \cite[Theorem 3.4]{bill_stability_2016} gives the existence of vectors $v, w \in \mR^K$ such that
\[ v^T A = r v^T, \quad A w = r w\,,\]
and $v$ and $w$ may chosen with every component positive. Further, by the same result it follows that
\[ \bM(t) e^{-rt} \to \frac{1}{v^T w} w v^T \bM(0) \quad \text{as $t \to \infty$}\,.\]
In particular, the limit on the left hand side of the above as $t \to \infty$ exists and is constant. Summing the components of $\bM$ gives that 
\[ \sum_{k=1}^K M_k(t) e^{-rt} \to \frac{1}{v^T w} \sum_{k=1}^K w_k \sum_{j=1}^K v_j M_j(0) \quad \text{as $t \to \infty$}\,.\]
Routine calculations give that the components of $w$ and $v$ respectively satisfy
\[ w_{k-1} = 2^{\frac{1}{K}} w_k \quad \text{for $k = 2, \dots, K$}\,,\]
and
\[ v_{k+1} = 2^{\frac{1}{K}} v_k \quad \text{for $k = 1, \dots, K-1$}\,.\]
If we take $w_K = 1$ and $v_1 = 1$, then, in light of the above,
\[ v^T w = \sum_{j=1}^K w_j v_j = K 2^{\frac{K-1}{K}}\,.\]
Moreover, as the components of $w$ comprise a geometric series with first term equal to one, we have that
\[ \sum_{k=1}^K w_k = \frac{(2^\frac{1}{K})^K-1}{2^{\frac{1}{K}} -1} = \frac{1}{2^{\frac{1}{K}} -1}\,.\]
Thus
\[ \lim_{t\rightarrow \infty} \sum_{k=1}^K M_k(t) e^{-rt} = \frac{1}{K 2^{\frac{K-1}{K}}(2^{\frac{1}{K}} -1)} \sum_{j=1}^K 2^{\frac{j-1}{K}} M_j(0),\]
as required.

\bibliographystyle{ieeetr}
\bibliography{main}

\begin{thebibliography}{10}

\bibitem{bremer_modulation_2008}
H.~Bremer and P.~P. Dennis, ``Modulation of chemical composition and other
  parameters of the cell at different exponential growth rates,'' {\em EcoSal
  Plus}, vol.~3, no.~1, 2008.

\bibitem{lindh_age_1999}
T.~Lindh and B.~Malmberg, ``Age structure effects and growth in the {OECD},
  1950–1990,'' {\em J Popul Econ}, vol.~12, no.~3, pp.~431--449, 1999.

\bibitem{gavagnin_invasion_2019}
E.~Gavagnin, M.~J. Ford, R.~L. Mort, T.~Rogers, and C.~A. Yates, ``The invasion
  speed of cell migration models with realistic cell cycle time
  distributions,'' {\em J Theor Biol}, vol.~481, pp.~91--99, 2019.

\bibitem{gabriel_contribution_2012}
P.~Gabriel, S.~P. Garbett, V.~Quaranta, D.~R. Tyson, and G.~F. Webb, ``The
  contribution of age structure to cell population responses to targeted
  therapeutics,'' {\em J Theor Biol}, vol.~311, pp.~19--27, 2012.

\bibitem{iwata_dynamical_2000}
K.~Iwata, K.~Kawasaki, and N.~Shigesada, ``A {Dynamical} {Model} for the
  {Growth} and {Size} {Distribution} of {Multiple} {Metastatic} {Tumors},''
  {\em J Theor Biol}, vol.~203, pp.~177--186, Mar. 2000.

\bibitem{ryser_quantifying_2016}
M.~D. Ryser, W.~T. Lee, N.~E. Ready, K.~Z. Leder, and J.~Foo, ``Quantifying the
  dynamics of field cancerization in tobacco-related head and neck cancer: a
  multiscale modeling approach,'' {\em Cancer Res}, vol.~76, no.~24,
  pp.~7078--7088, 2016.
\newblock Publisher: AACR.

\bibitem{castellanos-moreno_stochastic_2014}
A.~Castellanos-Moreno, A.~Castellanos-Jaramillo, A.~Corella-Madueño,
  S.~Gutiérrez-López, and R.~Rosas-Burgos, ``Stochastic model for computer
  simulation of the number of cancer cells and lymphocytes in homogeneous
  sections of cancer tumors.'' arXiv:1410.3768, 2014.

\bibitem{baar_stochastic_2016}
M.~Baar, L.~Coquille, H.~Mayer, M.~Hölzel, M.~Rogava, T.~Tüting, and
  A.~Bovier, ``A stochastic model for immunotherapy of cancer,'' {\em Sci Rep},
  vol.~6, no.~1, 2016.

\bibitem{gillespie_exact_1977}
D.~T. Gillespie, ``Exact stochastic simulation of coupled chemical reactions,''
  {\em The journal of physical chemistry}, vol.~81, no.~25, pp.~2340--2361,
  1977.
\newblock ISBN: 0022-3654 Publisher: ACS Publications.

\bibitem{gibson_efficient_2000}
M.~A. Gibson and J.~Bruck, ``Efficient {Exact} {Stochastic} {Simulation} of
  {Chemical} {Systems} with {Many} {Species} and {Many} {Channels},'' {\em J
  Phys Chem A}, vol.~104, no.~9, pp.~1876--1889, 2000.

\bibitem{yates_multi-stage_2017}
C.~A. Yates, M.~J. Ford, and R.~L. Mort, ``A {Multi}-stage {Representation} of
  {Cell} {Proliferation} as a {Markov} {Process},'' {\em Bull Math Biol},
  vol.~79, no.~12, pp.~2905--2928, 2017.

\bibitem{golubev_applications_2016}
A.~Golubev, ``Applications and implications of the exponentially modified gamma
  distribution as a model for time variabilities related to cell proliferation
  and gene expression,'' {\em J Theor Biol}, vol.~393, pp.~203--217, Mar. 2016.

\bibitem{chao_evidence_2019}
H.~X. Chao, R.~I. Fakhreddin, H.~K. Shimerov, K.~M. Kedziora, R.~J. Kumar,
  J.~Perez, J.~C. Limas, G.~D. Grant, J.~G. Cook, G.~P. Gupta, and J.~E.
  Purvis, ``Evidence that the human cell cycle is a series of uncoupled,
  memoryless phases,'' {\em Mol Syst Biol}, vol.~15, p.~e8604, Mar. 2019.

\bibitem{cao_analytical_2020}
Z.~Cao and R.~Grima, ``Analytical distributions for detailed models of
  stochastic gene expression in eukaryotic cells,'' {\em Proc Natl Acad Sci
  USA}, vol.~117, pp.~4682--4692, Mar. 2020.

\bibitem{jia_cell_2021}
C.~Jia, A.~Singh, and R.~Grima, ``Cell size distribution of lineage data:
  analytic results and parameter inference,'' {\em iScience}, vol.~24, no.~3,
  p.~102220, 2021.

\bibitem{mckendrick_applications_1925}
A.~G. McKendrick, ``Applications of {Mathematics} to {Medical} {Problems},''
  {\em Proceedings of the Edinburgh Mathematical Society}, vol.~44,
  pp.~98--130, 1925.

\bibitem{von_foerster_remarks_1959}
H.~von Foerster, ``Some remarks on changing populations,'' in {\em The
  {Kinetics} of {Cellular} {Proliferation}}, pp.~382--407, Grune and Stratton,
  1959.

\bibitem{arino_comparison_1993}
O.~Arino and M.~Kimmel, ``Comparison of {Approaches} to {Modeling} of {Cell}
  {Population} {Dynamics},'' {\em SIAM Journal on Applied Mathematics},
  vol.~53, pp.~1480--1504, Oct. 1993.

\bibitem{arino_survey_1997}
O.~Arino and E.~Sánchez, ``A {Survey} of {Cell} {Population} {Dynamics},''
  {\em Journal of Theoretical Medicine}, vol.~1, no.~1, pp.~35--51, 1997.

\bibitem{xia_pde_2020}
M.~Xia, C.~D. Greenman, and T.~Chou, ``{PDE} models of adder mechanisms in
  cellular proliferation,'' {\em SIAM Journal on Applied Mathematics}, vol.~80,
  no.~3, pp.~1307--1335, 2020.

\bibitem{greenman_kinetic_2016}
C.~D. Greenman and T.~Chou, ``A kinetic theory for age-structured stochastic
  birth-death processes,'' {\em Phys Rev E}, vol.~93, no.~1, p.~012112, 2016.

\bibitem{chou_hierarchical_2016}
T.~Chou and C.~D. Greenman, ``A {Hierarchical} {Kinetic} {Theory} of {Birth},
  {Death} and {Fission} in {Age}-{Structured} {Interacting} {Populations},''
  {\em J Stat Phys}, vol.~164, pp.~49--76, July 2016.

\bibitem{xia_kinetic_2021}
M.~Xia and T.~Chou, ``Kinetic theory for structured populations: application to
  stochastic sizer-timer models of cell proliferation,'' {\em Journal of
  Physics A: Mathematical and Theoretical}, vol.~54, p.~385601, Sept. 2021.

\bibitem{cao_accuracy_2006}
Y.~Cao and L.~Petzold, ``Accuracy limitations and the measurement of errors in
  the stochastic simulation of chemically reacting systems,'' {\em J Comp
  Phys}, vol.~212, no.~1, pp.~6--24, 2006.

\bibitem{yates_pseudo-compartment_2015}
C.~A. Yates and M.~B. Flegg, ``The pseudo-compartment method for coupling
  partial differential equation and compartment-based models of diffusion,''
  {\em J R Soc Interface}, vol.~12, no.~106, p.~20150141, 2015.

\bibitem{spill_hybrid_2015}
F.~Spill, P.~Guerrero, T.~Alarcon, P.~K. Maini, and H.~Byrne, ``Hybrid
  approaches for multiple-species stochastic reaction–diffusion models,''
  {\em J Comp Phys}, vol.~299, pp.~429--445, 2015.

\bibitem{erban_reactive_2007}
R.~Erban and S.~J. Chapman, ``Reactive boundary conditions for stochastic
  simulations of reaction-diffusion processes,'' {\em Phys Biol}, vol.~4,
  no.~1, pp.~16--28, 2007.

\bibitem{langlais_large_1988}
M.~Langlais, ``Large time behavior in a nonlinear age-dependent population
  dynamics problem with spatial diffusion,'' {\em J Math Biol}, vol.~26, no.~3,
  pp.~319--346, 1988.

\bibitem{torisawa_proliferation_2004}
Y.~S. Torisawa, H.~Shiku, S.~Kasai, M.~Nishizawa, and T.~Matsue,
  ``Proliferation assay on a silicon chip applicable for tumors extirpated from
  mammalians,'' {\em Int J Cancer}, vol.~109, no.~2, pp.~302--308, 2004.

\bibitem{rodriguez_wound-healing_2005}
L.~G. Rodriguez, X.~Wu, and J.-L. Guan, ``Wound-{Healing} {Assay},'' in {\em
  Cell {Migration}} (J.-L. Guan, ed.), pp.~23--29, Humana Press, 2005.

\bibitem{huyck_xtt_2012}
L.~Huyck, C.~Ampe, and M.~Van~Troys, ``The {XTT} cell proliferation assay
  applied to cell layers embedded in three-dimensional matrix,'' {\em Assay
  Drug Dev Technol}, vol.~10, no.~4, pp.~382--392, 2012.

\bibitem{vultur_cell--cell_2004}
A.~Vultur, J.~Cao, R.~Arulanandam, J.~Turkson, R.~Jove, P.~Greer, A.~Craig,
  B.~Elliott, and L.~Raptis, ``Cell-to-cell adhesion modulates {Stat3} activity
  in normal and breast carcinoma cells,'' {\em Oncogene}, vol.~23, no.~15,
  pp.~2600--2616, 2004.

\bibitem{simpson_migration_2010}
M.~J. Simpson, C.~Towne, D.~S. McElwain, and Z.~Upton, ``Migration of breast
  cancer cells: understanding the roles of volume exclusion and cell-to-cell
  adhesion,'' {\em Phys Rev E}, vol.~82, no.~4, p.~041901, 2010.

\bibitem{mcgary_cellular_2002}
E.~C. McGary, D.~C. Lev, and M.~Bar-Eli, ``Cellular adhesion pathways and
  metastatic potential of human melanoma,'' {\em Cancer Biol Ther}, vol.~1,
  no.~5, pp.~459--465, 2002.

\bibitem{garrido_hsp27_1997}
C.~Garrido, P.~Ottavi, A.~Fromentin, A.~Hammann, A.~P. Arrigo, B.~Chauffert,
  and P.~Mehlen, ``{HSP27} as a mediator of confluence-dependent resistance to
  cell death induced by anticancer drugs,'' {\em Cancer Res}, vol.~57, no.~13,
  pp.~2661--2667, 1997.

\bibitem{ratajczak_membrane-derived_2006}
J.~Ratajczak, M.~Wysoczynski, F.~Hayek, A.~Janowska-Wieczorek, and M.~Z.
  Ratajczak, ``Membrane-derived microvesicles: important and underappreciated
  mediators of cell-to-cell communication,'' {\em Leukemia}, vol.~20, no.~9,
  pp.~1487--1495, 2006.

\bibitem{davies_involvement_1998}
D.~G. Davies, M.~R. Parsek, J.~P. Pearson, B.~H. Iglewski, J.~W. Costerton, and
  E.~P. Greenberg, ``The {Involvement} of {Cell}-to-{Cell} {Signals} in the
  {Development} of a {Bacterial} {Biofilm},'' {\em Science}, vol.~280,
  no.~5361, pp.~295--298, 1998.

\bibitem{zakrzewski_stem_2019}
W.~Zakrzewski, M.~Dobrzyński, M.~Szymonowicz, and Z.~Rybak, ``Stem cells:
  past, present, and future,'' {\em Stem Cell Res Ther}, vol.~10, no.~1, p.~68,
  2019.

\bibitem{cassidy_role_2021}
T.~Cassidy, D.~Nichol, M.~Robertson-Tessi, M.~Craig, and A.~R.~A. Anderson,
  ``The role of memory in non-genetic inheritance and its impact on cancer
  treatment resistance,'' {\em PLOS Computational Biology}, vol.~17,
  p.~e1009348, Aug. 2021.

\bibitem{bill_stability_2016}
A.~Bill, C.~Guiver, H.~Logemann, and S.~Townley, ``Stability of {Nonnegative}
  {Lur}'e {Systems},'' {\em SIAM J Control Optim}, vol.~54, no.~3,
  pp.~1176--1211, 2016.

\end{thebibliography}

\end{document}